\documentclass[a4paper,11pt]{article}
\usepackage{jinstpub} % for details on the use of the package, please see the JINST-author-manual

\usepackage{graphicx}
\usepackage{subcaption}
\graphicspath{{figs/}} %Setting the graphicspath
\usepackage{siunitx}

\usepackage{listings}
\usepackage{xcolor}

\usepackage{lineno}
\title{\boldmath FELIX-MROD, a FELIX-based data acquisition system for the ATLAS Muon Drift Tubes}

\author[b,c,1]{Evelin Bakos\note{Now at: Alliander, Utrechtseweg 68, 6812 AH Arnhem, the Netherlands.}}
\author[a]{Henk Boterenbrood}
\author[b]{Mark Dönszelmann}
\author[e]{Florian Egli}
\author[b,2]{Luca Franco\note{Corresponding author.}} 
\author[b,3]{Carlo A. Gottardo\note{Now at: CERN, Esplanade des Particules 1, CH-1211 Meyrin, Switzerland.}} 
\author[b,4]{René Habraken\note{Now at: ASML Veldhoven, De Run 6501, 5504 DR Veldhoven, the Netherlands.}}
\author[b]{Adriaan König}
\author[a,d]{Antonio Pellegrino}
\author[e]{Chrysostomos Valderanis}
\author[a]{Jos Vermeulen}
\author[b]{Thei Wijnen}
\author[b]{Mengqing Wu}

\affiliation[a]{Nikhef, Science Park 105, 1098 XG Amsterdam, the Netherlands}
\affiliation[b]{Radboud University and Nikhef, Heyendaalseweg 135, 6525 AJ Nĳmegen, the Netherlands}
\affiliation[c]{Institute of Physics Belgrade, Pregrevica 118, Beograd, Serbia}
\affiliation[d]{Van Swinderen Institute, University of Groningen, Nijenborgh 3, 9747 AG Groningen, the Netherlands}
\affiliation[e]{Ludwig-Maximilians-Universitaet, Munich, Germany}

\emailAdd{l.franco@science.ru.nl}

\abstract{The ATLAS Muon Drift Tube (MDT) ReadOut Drivers (MROD), 204 VME modules that are an essential part of the readout chain
of the 1,150 MDT chambers, have been in operation for more than 15 years and are expected to remain in operation until about 2026.
In the event of extensive failures the number of spare MROD modules may be insufficient.  However, deployment of an adapted version
of the Front-End LInk eXchange (FELIX) system, a new component of the ATLAS data acquisition (DAQ) infrastructure, may overcome
potential MROD failures. This paper describes the design, functionality and performance of this adapted version, referred to as FELIX-MROD, and the test results of its integration into the ATLAS DAQ system}

\begin{document}
\maketitle
\flushbottom

\section{Introduction}\label{sec:intro}

This document describes the FELIX-MROD data acquisition (DAQ) system for ATLAS Muon Drift Tube (MDT) chambers~\cite{PERF-2007-01}. FELIX-MROD, that is based on the Front-End LInk eXchange (FELIX)~\cite{felixRef} and SoftWare ROD (SWROD)~\cite{swrodRef} systems, provides an alternative to the legacy readout architecture used in ATLAS since Run 1.

This document is organised as follows. The remainder of this section reviews the
current MDT readout chain and introduces the main components of the FELIX-MROD project. Section~\ref{sec:mrod} describes
the MDT ReadOut Driver (MROD) functionality and data structures. Section~\ref{sec:firmware}
is dedicated to the FELIX-MROD firmware, Section~\ref{sec:software} to the
FELIX-MROD software. Sections ~\ref{sec:commissioning} and~\ref{sec:integration} report on the testing
of FELIX-MROD in various test setups.
Conclusions are drawn in Section~\ref{sec:conclusion}.

\subsection{MROD: legacy readout of ATLAS Muon Drift Tube chambers}
The MDT chambers are the main component of the precision
tracking system in the ATLAS muon spectrometer~\cite{PERF-2007-01}.
An MDT chamber consists of assemblies of drift tubes of diameter \SI{29.97}{mm} filled
with $\text{Ar/CO}_{2}$. 

%[Three layers of MDT chambers surround the hadronic
%calorimeter in the barrel, and other three layers form the endcaps, covering up
%to $|\eta|<2.7$. In both the barrel and the endcap regions the layers are dubbed
%``Inner'', ``Middle'' and ``Outer''.]

The electrical signals generated by ionising particles crossing the tubes, are
routed to on-detector mezzanines once decoupled from high continuous voltage.
Each mezzanine can serve up to 24 drift tubes and is equipped with three
custom-designed monolithic Amplifier/Shaper/Discriminator (ASD) chips and one
Time-to-Digital Converter (TDC) chip~\cite{Arai:2008zzb}.
The ASD transforms the raw input from a tube into a pulse whose length is
proportional to the charge. The TDC digitizes the ASD signals and associates
them to a Level-1 accept signal (L1A) coming from the ATLAS trigger system:
buffered hits are associated to a L1A signal according to a configurable time
window. TDC output signals are transmitted via twisted pair cables to Chamber
Service Modules (CSMs).

A CSM~\cite{Arai:2008zzb} is a board
that incorporates one FPGA, one optical receiver for ATLAS trigger, timing, and control (TTC) signals, one gigabit optical
encoder chip and a VCSEL laser diode optical transmitter. One CSM can serve up
18 TDCs and acts as a multiplexer that serialises the electric TDC data streams
into a single optical stream run over a Gigabit Optical Link (GOL)~\cite{GOL}. 
The multiplexer operates as a simple rotating multiplexer that is continuously
checking each input TDC link for data present.
In addition, a CSM receives and delivers the TTC signals to all the mezzanines. The CSM board also interfaces to an Embedded Local Monitor Board (ELMB)~\cite{ELMB}, in the form of a so-called MDT-DCS module (MDM), used to communicate with the ATLAS
Detector Control System via CAN Bus.
A GOL link connects a CSM to an MDT ReadOut Driver (MROD) hosted in a
counting room (called USA15) next to the experimental cavern.

The GOL runs at 2 Gbit/s, using a low-level protocol in which 32-bit words are sent in a fixed length cycle of at maximum 21 8B/10B coded words. The first word is a cycle start word (``Separator word'') that has a fixed bit pattern (\texttt{0xd0000000} before 8B/10B coding). The Separator word is required to identify the data of the first TDC (number 0). It is followed by 18 words, one for each TDC. These cycles continue until all of the TDCs have delivered their data to the CSM. Each of these words is either a word coming from the TDC or an ``Empty word'' indicating that the TDC has no more data. The last two words in the cycle are encoded as 8B/10B comma patterns (IDLE words) to ensure the byte and 32-bit word alignment of the data on the optical link.

An MROD~\cite{Boterenbrood:2006cn, Arai:2008zzb} is a 9U VME module whose task is building
event fragments from the incoming time division multiplexed sequences sent by the CSMs,
reporting and correcting errors in the data and reducing the data size.
One MROD module can receive the data streams from up to six CSMs, is interfaced
to the TTC system and the output is transferred via a ReadOut Link implemented as an S-Link~\cite{SLINK}.
On the other end, the S-Link is connected to the ATLAS Trigger and Data
Acquisition (TDAQ) System. More specifically, the S-Link is connected to a
RobinNP PCIe board hosted on a ReadOut System (ROS) server~\cite{ROS, Borga_2023}.
The RobinNPs and therefore also the ROS servers are not detector specific. The task of a ROS server
is to aggregate data coming from up to 24 S-Links and transmit that data on demand to
the High Level Trigger (HLT) via commercial 10 Gigabit Ethernet links.
A sketch of this \textit{legacy} readout architecture is shown in the left part of \figurename~\ref{fig:mergerMROD}.

MROD modules have originally been produced and installed in 2006. 
By the end of Run 3, now expected for 2026, most modules will have been in continuous operation for twenty years.
It is difficult to estimate the lifespan of an MROD module. So far, of 204 MROD modules, 26 had to be replaced and at the end of 2022 the number of MROD spares was 21.
In case of extensive failures, or damage to an entire crate or rack, the number of spare modules might be insufficient.
New modules cannot be produced due to the obsolescence of components and a different solution would be necessary.
In such exceptional case, FELIX-MROD could be deployed.
FELIX-MROD makes use of the two new components of the recently upgraded ATLAS DAQ system: FELIX and SoftWare ROD (SWROD).

\subsection{Readout architecture based on FELIX and SWROD}
During the second long shutdown (LS2), the ATLAS detector was upgraded with new sub-detectors (NSW, BIS7/8) and an improved TDAQ system~\cite{Izzo:2732959}.
The trigger system was enhanced by increasing the readout granularity of the
Liquid Argon (LAr) calorimeter and including feature extractors that target specific objects.
The DAQ system was also upgraded with new components, the new FELIX and SWROD, in order to support the new sub-detector and trigger systems~\cite{Gottardo:2744566}.

A FELIX unit consists of one server equipped with a multi-Gigabit Ethernet adapter and one or two custom PCIe cards, called FLX-712, suitable for all sub-detectors.
A single FLX-712 card can serve up to 48 channels (thus corresponding to 8 MRODs) and can handle messages from the L1 trigger.
FELIX acts only as a router that forwards incoming data, without any alteration, to the SWROD units over an Ethernet network.
A SWROD unit, equipped with a dedicated plug-in application to deal with CSM data, consists of a commodity server running software that performs data processing and aggregation.
SWROD servers are interfaced with HLT and provide it with data information on demand, similarly to what is done by the ROS component of the legacy architecture.

\begin{figure}%[htbp!]
    \centering
    \includegraphics[width=\textwidth]{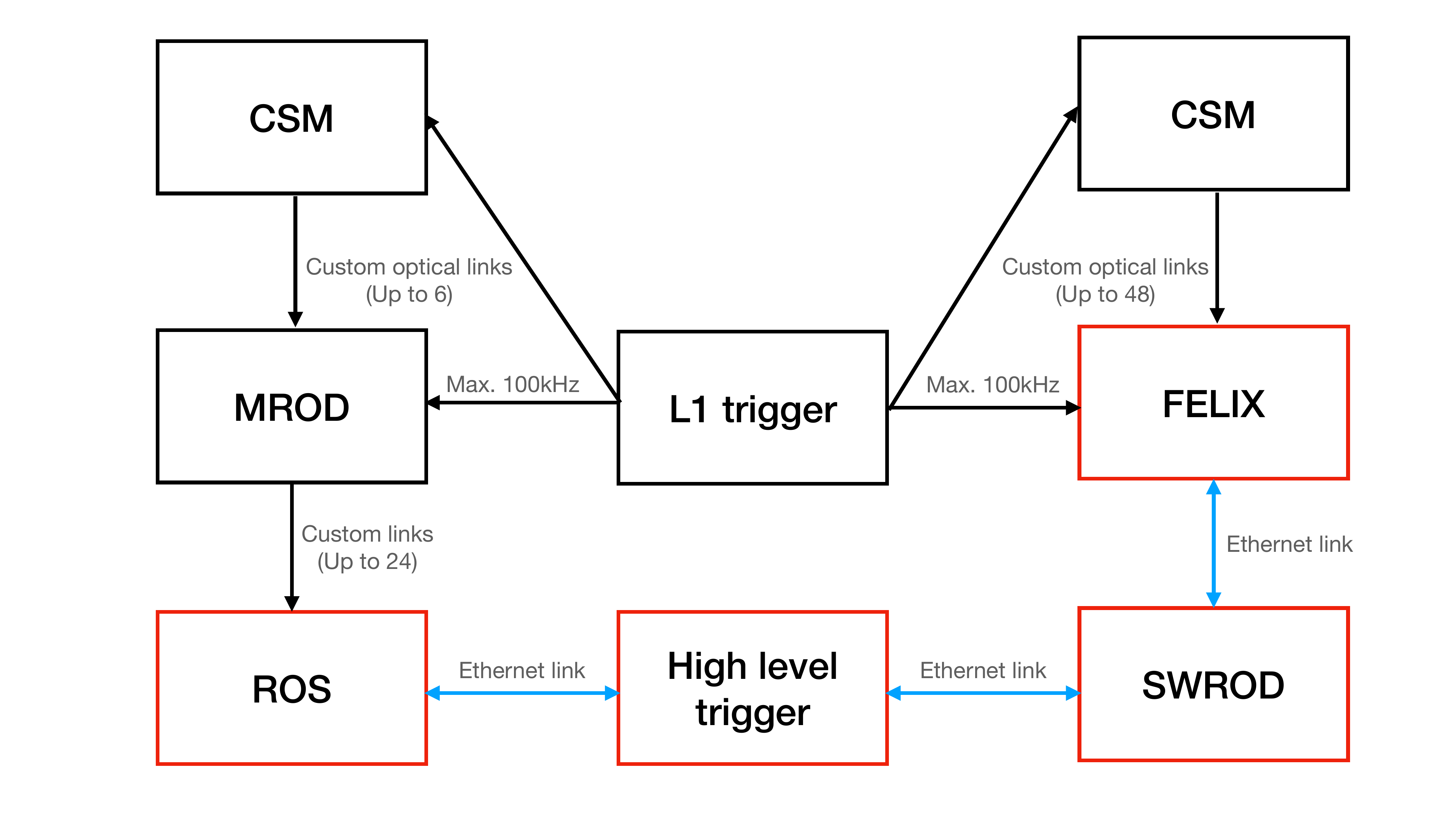}
    \caption{Schematic representation of the legacy readout architecture next to the alternative one where FELIX-MROD is introduced.}
    \label{fig:mergerMROD}
\end{figure}

As shown in \figurename~\ref{fig:mergerMROD}, the only custom
component in the FELIX-MROD architecture is in fact the FELIX PCIe
card, universal for all the detector systems. Having decoupled
the data transport and processing, and making use of industry
standard data networks earlier in the readout chain, the new
architecture is also more modular and flexible and has a much lower
hardware and firmware development and maintenance cost
than the legacy one.

During Run 3, ATLAS uses 60 FELIX
servers, 96 FELIX cards and 30 SWROD servers. In the High-Luminosity phase of LHC,
FELIX will be adopted by all ATLAS sub-detectors.

\section{Functionality of the MROD}\label{sec:mrod}
\begin{figure}[htbp]
    \centering
    \includegraphics[angle=90, width=\textwidth]{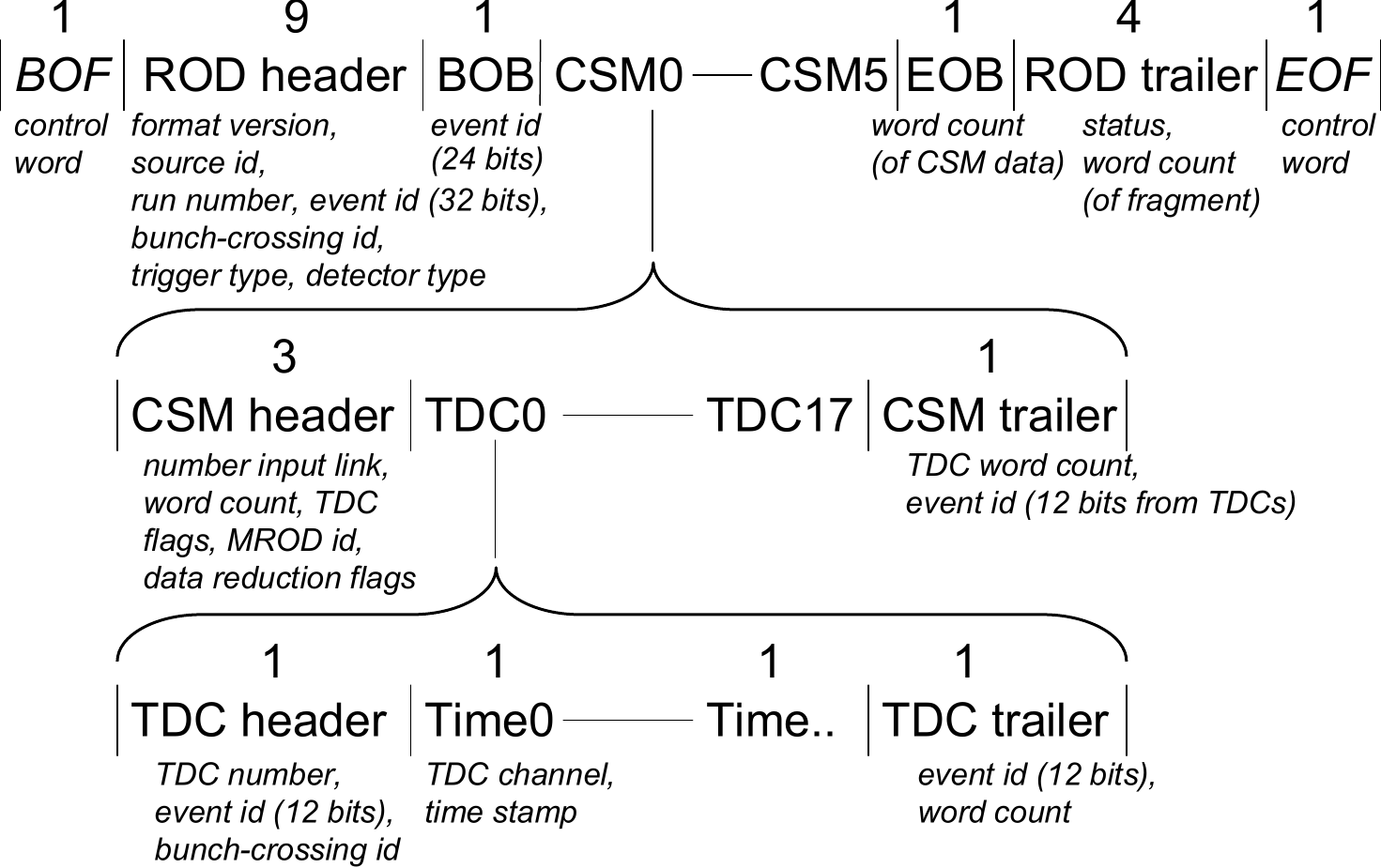}
    \caption{Structure of the MROD output data format.~The numbers indicate the maximum number of data-words of a given type.}
    \label{fig:mrod_data_format}
\end{figure}

The main task of the MROD is to demultiplex data streams coming from
multiple CSMs, and build event fragments to be sent to the ROS.
The MROD data output in structured in three nested levels  
\cite{Wijnen:2003caa} as represented in Figure~\ref{fig:mrod_data_format}.
The lowest level is the TDC level. The next upper level, the CSM level, is a
collection of the TDC fragments, while the highest level corresponds to a group
of up to 6 chambers defining the MROD level fragment. Each level envelopes
the previous one and has the same basic structure: one or more header words
followed by a number of data words and terminated with a trailer containing the
word count for the fragment as a whole. Fragments can be empty, i.e. not containing data. At each level the fragments of the next lower level are fully
included in the fragment of the current level.  The TDC envelopes are generated
by the TDCs in the form of header and trailer words. In the absence of hits, a
TDC sends an empty envelope for each L1 trigger to the CSM. Since the CSM is
virtually transparent to the data, the CSM level envelope is generated by the
MROD (MROD specific information is also added), as is the MROD level envelope.
Empty TDC envelopes can be discarded by the MROD to reduce the size of the event
fragments output. 

Every event constructed by an MROD has to fulfill the ATLAS TDAQ event format
requirements~\cite{Bee:683741}. The full event is built from sub-detector
fragments, where each of these fragments is an aggregation of ROS fragments. 
Each ROS fragment is an aggregation of ROB fragments, which contain one ROD fragment. The format of ROS, ROB and ROD headers is the same for all
ATLAS sub-detectors and all headers begin with a Start of Header Marker.

In case of the MDT chambers, the general ROD header and trailer words
are constructed by the MROD. These words are the following:

%###\setlist{nolistsep}
%### removed [noitemsep]
%###\begin{itemize}[noitemsep]
\begin{itemize}
\item ROD header words (each word is 32-bit):
%###\begin{itemize}[noitemsep]
\begin{itemize}
\item \textbf{Start of ROD Header marker}: indicates the
start of a ROD fragment header, which is defined to be \texttt{0xEE1234EE} for
all RODs.
\item \textbf{ROD Header size}: indicates the size of the header including the
header marker. The header contains 9 words.
\item \textbf{Format version number}: indicates the format version of this ROD
fragment. The upper two byte state the major and the lower two
byte state the minor version number.
\item \textbf{Source identifier}: identifies the origin of the ROD fragment.
The word consists of a module type, a sub-detector ID and a module ID. The
sub-detector IDs for the MDT detector are in the range  \texttt{0x61} through 
\texttt{0x64}. 
\item \textbf{Run number}: the highest 8 bits are defined by the type of the
Run (physics, calibration, etc.). The low order 24 bits represent the ordered
sequence of runs within a type.
\item \textbf{L1ID}: contains the 32-bit event identifier generated by the L1
trigger system.
\item \textbf{BCID}: contains the 12-bit bunch crossing identifier
generated by the L1 trigger system.
\item \textbf{L1 Trigger type}: contains the 8-bit event trigger type
defined by the L1 trigger system.
\item \textbf{Detector event type}: identifies an event which may have been
generated by a sub-detector, independently of the ATLAS trigger systems.
\end{itemize}
\item ROD fragment trailer types:
%###\begin{itemize}[noitemsep]
\begin{itemize}
\item \textbf{MROD status element (MSE)}: when zero it shows that no known
errors are associated with this event fragment.
\item \textbf{Number of status elements (NSE)}: total number of words that were
inserted in the status block. At least one status
word must be present in the event.
\item \textbf{Number of data elements (NDE)}: total number of words in the data
block, excluding the 9 words in the MROD header.
\item \textbf{Status block position (SBP)}: defines the relative order
of the data and status elements. A value of zero indicates that the
status block is placed before the data block and a value of one indicates
that the status block is placed after the data block.
\end{itemize}
\end{itemize}

Besides the general ROD header and trailer words, additional, MROD-specific
header and trailer words are also included to the data stream. The MROD specific header and trailer words are the following:
%###\setlist{nolistsep}
%###\begin{itemize}[noitemsep]
\begin{itemize}
\item \textbf{MROD header, BOB (MROD Begin Of Block)}: The BOB word, bit pattern \texttt{0x80nnnnnn}, the lower 24 bits are used to store a copy of the L1ID.
\item \textbf{MROD trailer, EOB (MROD End Of Block)}: The EOB word, bit pattern
\texttt{0xF0nnnnnn}, contains 16 bits of word count for this MROD event block,
counting from and including the BOB word, up to and including the EOB word.
\end{itemize}

The BOB and EOB surround the CSM data envelope that consists of several CSM fragments, each preceded by a header and followed by a trailer, as specified below.
%
%###\setlist{nolistsep}
%###\begin{itemize}[noitemsep]
\begin{itemize}
\item Header words:
%###\begin{itemize}[noitemsep]
\begin{itemize}
\item \textbf{MROD LWC (Link Word Count)}: the LWC word, bit pattern
\texttt{0x81nnnnnn}, indicates the first word of an event fragment
coming from one CSM. This word contains 4 bits of the L1ID for debugging
purposes and 16-bit word count for the number of words in this CSM
fragment including the LWC itself and the TWC word. The LWC word
is always followed by the BOL word.
\item \textbf{MROD BOL (Begin Of Link)}: the BOL word, bit pattern \texttt{0x18nnnnnn}, indicates which CSM link is associated with the data. It contains 4 bits for
the CSM link number, 12 bits for the MROD module serial number, and a
number of status bits indicating different working conditions. The
status bits report whether the TDC data is zero-suppressed, trailer
suppressed and the type of TDC (AMT or HPTDC).
\item \textbf{MROD TLP (TDC Link Present)}: the TLP word, bit pattern
\texttt{0x89nnnnnn}, contains status information about the TDC links
connected to this CSM, each bit in the 18 lower positions represents a TDC that has sent data for this event. This header is always present, 
even if this CSM had no data at all.
\end{itemize}
\item Trailer word:
%###\begin{itemize}[noitemsep]
\begin{itemize}
\item \textbf{TWC (Trailer Word Count)}: the TWC word, bit pattern
\texttt{0x8Annnnnn}, indicates the last word of an event fragment
coming from one CSM link. It contains 12 lower bits of the 24-bit L1ID
stored in the MROD. It also contains 12 bits of word count for the number
of words coming from this CSM link, starting from (and including) the
TLP word and all TDC words up to and including the TWC word itself.
\end{itemize}
\end{itemize}

The CSM aggregates data from up 18 TDCs, which means, that within the CSM event
fragment, there are up to 18 TDC event fragments. Every data fragment sent by a TDC contains the following header and trailer word:
%###\setlist{nolistsep}
%###\begin{itemize}[noitemsep]
\begin{itemize}
\item \textbf{TDC data header, BOT (Begin Of TDC)}: the 
TDC BOT word, bit pattern \texttt{0xAtnnnnnn} or \texttt{0xBtnnnnnn}, marks the start of TDC data  for this event. The header starts with 0xA for
words from TDC 0-15 and with 0xB for BOT words from TDC 16-17 and
$t=$ 0{\ldots}0xf encodes TDC number 0{\ldots}15. It also contains 12 bits for the event count generated by the TDC, and 12 bits of BCID. 
\item \textbf{TDC data trailer, EOT (End Of TDC)}: the TDC EOT word, bit pattern \texttt{0xCtnnnnnn} (where $t=$0), indicates the end of TDC data
fragment for this event. It contains 12 bits of event counter and also 12-bits
word count, which indicates the number of words in this TDC data block including
itself.
\end{itemize}

\begin{figure}[htbp]
    \centering
    \fbox{\includegraphics[width=\textwidth]{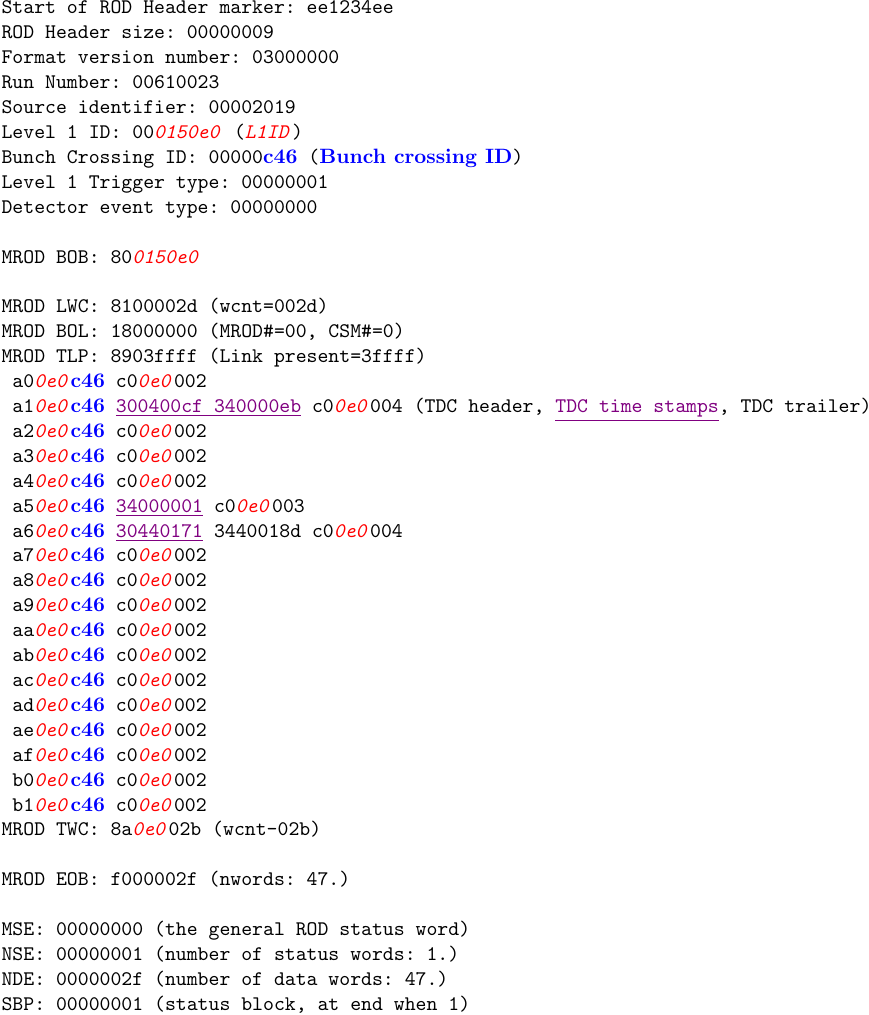}}
    \caption{An example of an MROD event, shown here in a human-readable form. The L1ID trigger number (or its lower 12-bits) are indicated in \textcolor{red}{\textit{red (italic)}}, the BCID number in \textcolor{blue}{\textbf{blue (bold)}} and the TDC time stamps in \textcolor{violet}{\underline{violet (underlined)}}.}
    \label{fig:mrodevent}
\end{figure}

For normal running conditions the presence of the following words in each event
is expected: the BOB word, followed by a maximum of 6 CSM blocks, each
consisting of the LWC, BOL and TLP words, a number of TDC words and a
terminating TWC word. The last CSM block is terminated by the EOB word.
This can be seen in the example event shown in Figure~\ref{fig:mrodevent}.
In this example event only one CSM block is present. When more CSMs are connected,
the block between the MROD LWC and MROD TWC word is present for each CSM.
CSM data is ordered according to TDC number (in this case from 18 TDCs in total),
surrounded by words, whose names start with ``MROD'', added by the 'MRODin' (except BOB and EOB),
and the outer ROD header and trailer sets of words (plus BOB/EOB), added by the 'MRODout'.
Data from individual TDCs start with the word that starts with 'a' or 'b' (Begin-Of-TDC word or BOT)
and end with the word that starts with 'c' (End-Of-TDC word or EOT), and only TDCs with hits for this event
have data words (words starting with '3') between the BOT and EOT.
The numbers in the TDC BOT and EOT words
have to correspond to the corresponding values in the headers
(which are derived from information received from the trigger system during data taking),
to ensure the data-acquisition is synchronized.

The MROD will optionally perform zero suppression. To allow consistency checks
at the input of all six CSM links, each TDC will always send both header and
trailer words, even if there is no real measurement data present. The MROD can check
for each TDC whether it only received a BOT and EOT word with the correct event
ID encoded in both words. If that is the case, both words may be skipped when sending data to the MROD output link. In the case where
there is real data coming from a TDC, the BOT cannot be suppressed because it
is the only word with the full TDC identifier. When trailer-suppression is
turned on, the TDC trailer word is skipped even if that TDC contained hit data, but only if it contained a correct event number and a correct word count. The EOT word will only be removed if it contained a correct event
number and a correct word count. If one of them is incorrect, the EOT will
still be placed in the data stream. The BOL word shows whether either one or
both suppression modes have been turned on.

\subsection{Event building mechanism}
Each MROD module uses Xilinx Virtex-II Pro FPGAs for data processing,
which can be divided into two categories, processing data from the CSMs by the \texttt{MRODin} FPGAs and processing of their output data by the \texttt{MRODout} FPGA. There is one \texttt{MRODin} FPGA per CSM link which, in the absence of errors, processes the CSM data without intervention.
The FPGA demultiplexes the CSM data, removes the
empty and Separator words, and reconstructs the data streams of the individual
TDCs. After reconstruction, data are subsequently stored in external memory associated with the FPGA. The FPGA recognizes the TDC trailer words and records L1 trigger
information. As described in the previous section, TDC trailer words and/or empty TDC envelopes may optionally be suppressed.
Once the FPGA has recognized a complete event and the trailer words of all 
active TDCs have been received, the CSM level header and trailer words are
generated. The \texttt{MRODin} passes the data on to the \texttt{MRODout}.
The first step of the fragment building in the \texttt{MRODout} FPGA consists
of generating the correct MROD header words. The MRODout FPGA sends the output
streams of the MRODin FPGAs one after the other to the MROD output link.
When all the CSM fragments have been sent out, the data stream is terminated with the
trailer words.

During run time, the FPGA checks for a number of error and exception conditions:
%###\setlist{nolistsep}
%###\begin{itemize}[noitemsep]
\begin{itemize}
    \item Parity errors on the TDC to CSM link (these errors are encoded in the data by the CSM),
    \item Link and/or parity errors on the CSM to MROD link
    \item Absence of data, incorrect or too long event fragments from a TDC
    \item Absence of expected trailer words or corruption of trailer words
\end{itemize}

In case an error is detected, the FPGA intervenes appropriately.
For very serious conditions (too large event fragment or a memory buffer overrun)
the FPGA will independently decide to ignore an individual TDC channel.

%\subsection{ATLAS Phase-I FELIX}

\section{FELIX-MROD firmware}\label{sec:firmware}
An FLX-712 card is equipped with all the necessary hardware to read out up to
48 CSM modules and transfer the incoming data to the memory of the host server.
In order to interpret the CSM data format and re-implement the MROD
functionalities in FELIX, custom FELIX-MROD firmware has been developed.
Each of the 48 inputs (2 x 24) of the FLX-712 card can handle
one GOL running at 2 Gbit/s. 

%%%%%%%%%%%%%%%%%

%[The MDT chambers are equipped with on-chamber electronics that process
%the signals from the drift tubes, connected to a TDC. There are at maximum 18 TDC's (each one has 24 inputs) on
%a chamber and they are read out by the CSM.
%The CSM and the TDCs are connected to the TTC system and they will
%start the readout on a L1A trigger signal. For each L1A, all
%of the TDCs will deliver two or more words to the CSM: a TDC header
%word, zero or more (up to 100) TDC timestamp words and a TDC trailer
%word.]

The 8B/10B coded 32-bit data received via the GOL are decoded at a rate of 50 MHz, comma patterns do not result in data output by the decoder. Depending on the data contents and the mode of operation, for each GOL all or part of the decoded data is stored in a FIFO by the ``CSM handler''. The modes of operation are:

\begin{enumerate}
    \item Pass-all: in this mode all data (Separator words, valid TDC data words or Empty words)  are pushed in the FIFO;
    \item Empty Cycle suppress mode: in this mode (default) all data is pushed as long as there is at least one valid TDC data word in the cycle. If all 18 TDC words appear to be an Empty word, the whole cycle is suppressed;
    \item Empty Suppress mode: in this mode, the Separator word is updated to hold bit flags for each valid TDC word. The lower 18 bits of the Separator word preceding the data hold the flags and it is followed by 18 or less TDC words. 
\end{enumerate}

The logic on the output side of the ``CSM handler'' FIFO accumulates the data and sends it to the Central Router of the FELIX firmware
in chunks of 254 words. The Central Router adds a block header and a chunk/block trailer and then
it pushes the 1 kByte block into the large ``toHost'' FIFO. The Central Router has two DMA endpoints, each handling 24 input channels.

The 1 kByte blocks arriving in the host are sent to the SWROD, which
handles the decoding and formatting of CSM data into MROD event
fragments, as described in detail in Section~\ref{sec:software}.

%The format of the data output by the MROD can be visualized as
%consisting of three nested levels of envelopes, as described in detail in Section~\ref{sec:mrod}.
%In the SWROD it is also possible to reduce the size of the event
%fragments by suppressing empty envelopes: TDCs that only send a header
%and a trailer word, may be suppressed.

\begin{figure}%[htbp!]
    \centering
    \includegraphics[width=\textwidth]{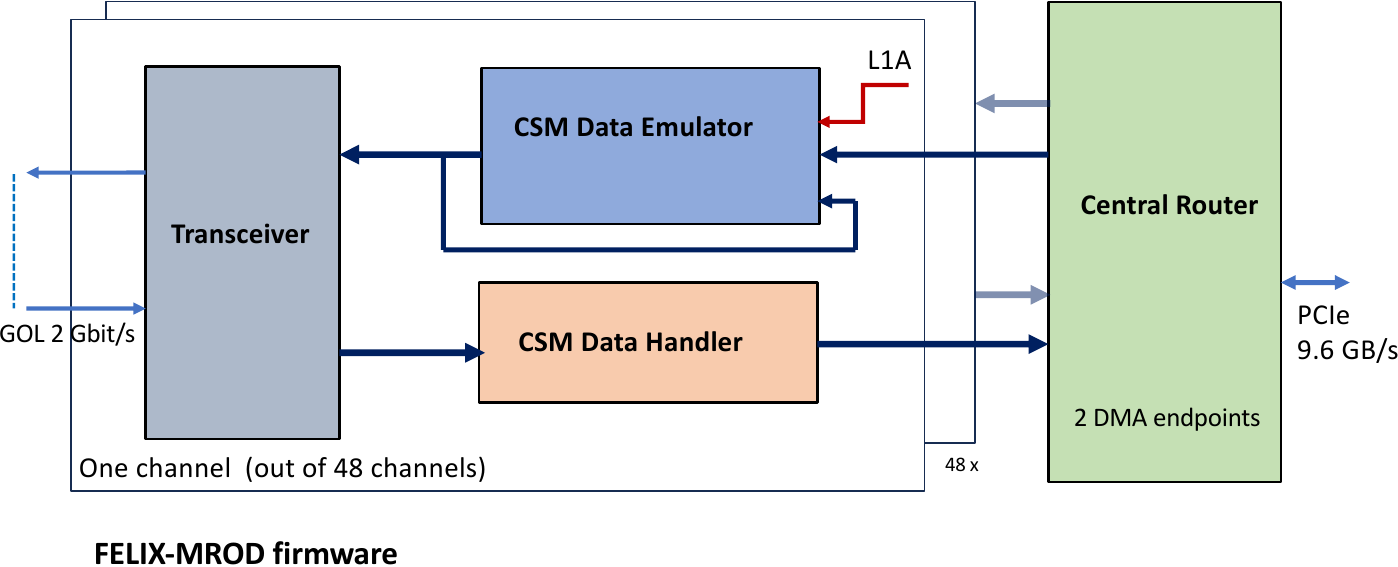}
    \caption{Block diagram representing the input (output) data stream from (to) the transceivers to (from) the host server.}
    \label{fig:firmwareSketch}
\end{figure}

A block diagram of the firmware design is given in \figurename~\ref{fig:firmwareSketch}.
The firmware features an emulator that enables thorough testing of the CSM Handler and the Central Router. Each channel has its own emulator that can be loaded from the host server with a data pattern that resembles real CSM data. It uses the L1A trigger signal to send out data from one event. At the same time, this data is looped back to the input, while on the fly adjusting the 12 bit L1ID in the TDC header and trailer words. The CSM Handler will thus see a continuously increasing L1ID which is what the SWROD expects to see.
The data can be looped back internally (without going through the transceivers) or externally by using a fiber loop from output to input.

%%%%%%%%%%%%%

\section{Software}\label{sec:software}
\subsection{FELIX software}\label{sec:fsoftware}

The primary data routing software infrastructure in charge of
transferring data to and from the FELIX host is called \texttt{felix-star}. It comprises a set of single-threaded applications: \texttt{felix-tohost} to read the data
coming from the front-end, \texttt{felix-toflx} to send data from the host to the
front-end, and \texttt{felix-register} to configure FELIX cards parameters remotely.
In the downstream direction, \texttt{felix-star} sends data to remote clients (i.e. SWROD units) according
to their link subscription. Active links published by \texttt{felix-star} are advertised
by the \texttt{felix-bus} system using a set of JSON files accessible by both server
and clients. Communication between \texttt{felix-star} applications and network peers
is managed by the \texttt{netio-next} library, based on \texttt{libfabric} and capable of
exploiting RDMA technology.
The mainstream FELIX software release supports the FELIX-MROD firmware and all necessary MROD-specific support features have been added to \texttt{felix-star}. 

\subsection{CSM-SWROD}\label{sec:CSM-SWROD}
FELIX firmware and software do not perform any data processing, but only
provide data routing between detector front-end and the DAQ system. The task of
data aggregation and processing is fulfilled by SWROD: an application running on commodity servers.
To accommodate the needs of multiple sub-detectors, SWROD has been designed to
support a high degree of customization. Its architecture comprises a set
of independent components, with each of them providing a simple
interface that defines how other components can interact with it. 
These interfaces are:
\begin{itemize}
%\item \texttt{DataInput} interface: receives data from the network. It protects
\item \texttt{DataInput} interface. It protects
the other components of the SWROD application from any changes in the network
input protocol, and gives a possibility to use another data source than the network input for testing and debugging.
\item \texttt{ROBFragmentBuilder} interface of the main data aggregation algorithm, which aggregates data chunks from individual e-links, received via the \texttt{DataInput} interface, into event fragments according to a given configuration. Such a configuration defines the set of event fragments to
be produced as well as a list of input links for each fragment.
\item \texttt{ROBFragmentConsumer} interface for the application of custom subdetector specific processing of the fully aggregated event fragments before these are passed to the HLT farm.
\end{itemize}

\begin{figure}[htbp!]
    \centering
    \includegraphics[width=\textwidth]{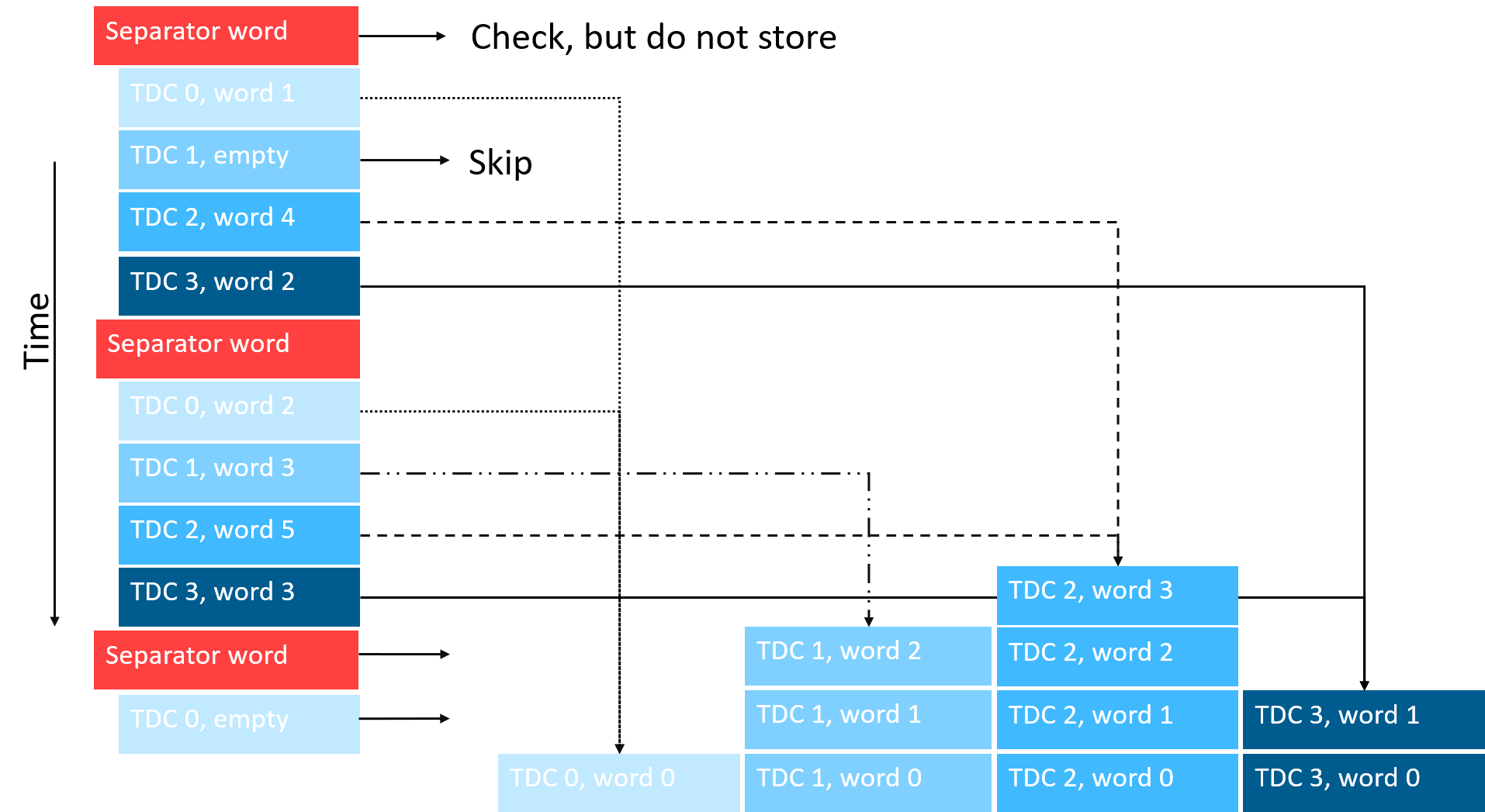}
    \caption{Representation of the \texttt{CSMProcessor} event building functionality, identical to that of the MROD. The left column indicates the incoming data, which is stored in separate buffers based on the TDC number.}
    \label{fig:fifo}
\end{figure}

The SWROD implementation that reproduces the MROD functionality is called
CSM-SWROD and includes a custom \texttt{ROBFragmentBuilder} called \texttt{CSMBuilder}, and a custom
processing plugin.
%CSMBuilder
The main task of \texttt{CSMBuilder} is to receive the 32-bit trigger identifier
called L1ID from the TTC link and to check it against the L1ID retrieved from data.
The task is performed by a set of \texttt{CSMWorker} threads, each equipped with a 
\texttt{CSMProcessor}. 
%CSMProcessor
The \texttt{CSMProcessor} receives data, strips off the Separator and the
``Empty words'', and sorts the incoming data into different buffers based on the
TDC number. The BOT and the EOT words are also stored in these buffers. \texttt{CSMProcessor} builds the event by sorting the
TDC data such that all TDC event fragments are starting with the BOT word
and terminated by the EOT word. When all TDC inputs delivered an EOT word 
\texttt{CSMProcessor} packs the TDC event fragments together. 
This event building procedure, shown in Figure \ref{fig:fifo}, is the same as that of the MROD.
After the event is built, \texttt{CSMProcessor} constructs the header and
trailer words corresponding to the CSM and MROD data packets and inserts them in their predefined places within the data fragment. To build the header and trailer
words additional information is needed about the MDT chambers.
These parameters need to be specified in the general \texttt{SWRODSegment}
configuration file, which contains also other configuration information needed for the 
SWROD operation. The CSM-SWROD specific settings are the following:
\begin{itemize}
\item \texttt{MRODSerialNumber}: contains the MROD serial number. This information is needed for the BOL header word.
\item \texttt{CSMInputNumber}: contains the CSM number corresponding to the
input link. This information also necessary for constructing the BOL header word.
\item \texttt{HPTDC}: a flag which determines the type of TDC used on this
chamber. This flag is stored in the BOL header word as well.
\item \texttt{ZeroSuppression} and \texttt{TrailerSuppression}: flags that
indicate if zero suppression or trailer suppression, see Section~\ref{sec:mrod} for a description of this functionality, is turned on.
These flags are also present in the BOL header word. 
\item \texttt{TDCChannelMask}: Indicated which TDCs are present for the given
input link. This information is also presented in the TLP header word.
\end{itemize}
With the parameters above, it is possible to build every header and trailer word
needed to construct the full ROD fragment, except the MROD header and the BOB word.
These contain the full L1ID, which cannot be determined from the TDC data. To be able to construct the BOB word, an additional plugin is needed: the \texttt{CSMCustomProcessor}, called via the \texttt{ROBFragmentConsumer}. It retrieves the full trigger information from the TTC input and stores the TriggerType, the BCID and the L1ID in the MROD header and the L1ID in the BOB header word.
The plug-in is also responsible for
configuration and addition of the status words in the end of the data fragment.

%\begin{figure}[htbp]
%\centering
%\includegraphics[width=\textwidth]{figures/csmSWROD.png}
%\caption{CSM-SWROD interface implementations. \texttt{CSMBuilder}
%is part of the \texttt{ROBFragmentBuilder} interface, while
%\texttt{CSMCustomProcessor} is part of the \texttt{CustomProcessor} interface.}
%\label{fig:csmSWROD}
%\end{figure}

In addition to the fragment building, the \texttt{CSMProcessor} is also validating
the data in each TDC fragment, as well as performing a synchronization procedure,
to make sure that the event fragments with the same L1ID are accepted from the
TDC links and are sent out together. At this step the \texttt{CSMProcessor} can
perform channel suppression in case one of the TDC channels falls out of
synchronization. In this case, if multiple channels have data but one or more is
lagging behind, the given channel is turned off, and the TLP word is updated
accordingly. The channel is checked on every L1ID rollover, and in case it is
supplying data again, it is turned back on. 
Another important part of the CSM-SWROD is the error reporting system,
which includes all the error checks implemented in the MROD. These errors are defined within the Error Reporting System (ERS) of the
SWROD and are reported alongside with the general SWROD warnings or
errors.  Besides the already defined error conditions, a new feature has also been added to the \texttt{CSMProcessor}, which was not part of the original MROD
system. The \texttt{CSMProcessor} can create fake BOT and EOT words if these are missing for given TDC channel, which ensures continuous data processing and building. Both the BOT and EOT word contain information which can be retrieved from the data itself, so it is possible to
create them without any intervention. In case this happens, a flag is set in
the corresponding EOT word, to indicate that the EOT or the BOT were
artificially introduced during data processing. When the complete event has been built, it is sent to the \texttt{CSMBuilder} and then to the
\texttt{ROBFragmentConsumer} interface for further processing.

\section{Commissioning}
\label{sec:commissioning}
% \subsection{Test with pre-recorded data} %on MDT, then on FELIX
% \subsection{Test with MDT chambers}

\subsection{Software test with pre-recorded data}
The first test of FELIX-MROD involved only CSM-SWROD and was performed without a
FELIX card. A file containing pre-recorded MDT data was provided to
\texttt{felixcore} (predecessor of \texttt{felix-star}) and events were sent to CSM-SWROD over the
network.
This approach allowed to simulate up to 10 CSMs in a realistic scenario
and evaluate the feasibility of implementing the fragment building algorithm in
software. CSM-SWROD was found to be able to process incoming messages within
\SI{8}{\micro\second} deploying one thread per CSM on an Intel Xeon Gold 5115
CPU. 
%The processing time distribution is shown in Figure~\ref{fig:swrod_proc_time}: input messages containing no hits are processed in shorter time (first peak) compared to messages with hits (second peak).
%
%\begin{figure}[htbp]
%    \centering
%    \includegraphics[width=0.5\textwidth]{figures/csmswrod_time.png}
%    \caption{Processing time of FELIX messages.}
%    \label{fig:swrod_proc_time}
%\end{figure}

%\subsection{Test with pre-recorded data loaded on FELIX}
%The functionality of the MROD firmware was tested with both FELIX low-%level tools
%(\texttt{fdaq}) and felix-star using a loopback mode implemented in the
%FELIX MROD firmware. CSMswROD could assemble fragments in this but the TTC %link
%could not be integrated due to the hard-coded L1IDs in data. 

\subsection{Performance tests at Radboud University}
\begin{figure}
    \centering
    \includegraphics[width=\textwidth]{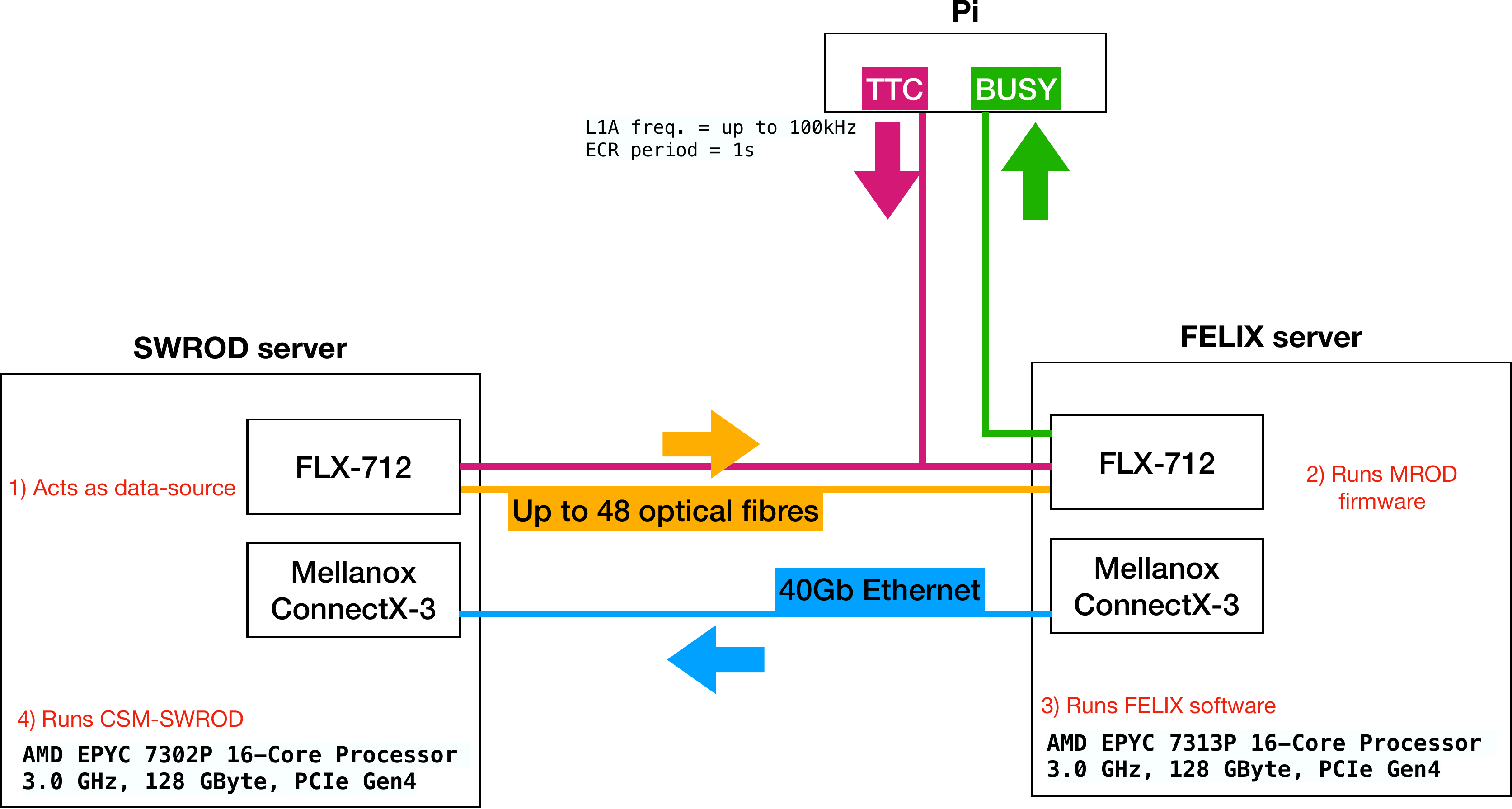}
    \caption{Schematic view of the setup used to perform the tests at Radboud.} 
    \label{fig:RUscheme}
\end{figure}

FELIX-MROD has been tested at the High Energy Physics department's laboratory of the Radboud University in Nijmegen, with another setup than the one used during the development stages. 
This setup is schematically represented in Figure~\ref{fig:RUscheme} and consists of two servers, the SWROD server and the FELIX server, both hosting a FLX-712 card
and equipped with a 40 GbE Mellanox ConnectX-3 NIC to allow communication over the Ethernet network.
The FELIX server has a 16-core AMD Epyc Milan 7313P CPU while the SWROD server has a 16-core AMD Epyc Rome 7302P CPU.
Both have 128 GB of memory and run the CentOS 7 operating system.
The TTC system is controlled by a Raspberry Pi and its output is connected to the mezzanines of both FLX-712 cards.
The described setup was exploited to carry out performance tests:
\begin{enumerate}
\item The FLX-712 card in the SWROD server, configured with special emulator firmware, acts as a data source;
\item The other FLX-712 card in the FELIX server reads out the data through the optical link, upon reception of L1 accept signals (L1A) generated by the TTC;
\item The FELIX software runs in the FELIX server and is responsible for routing the data from the card to the network;
\item The two servers communicate through a high-speed Ethernet network using 40 GbE Mellanox interfaces;
\item The CSM-SWROD application running in the SWROD server processes the incoming data from the network.
\end{enumerate}
The throughput of the card in the FELIX server was measured for stepwise increasing trigger rates: L1A signals were sent to both FLX-712 cards to trigger the data-emulation and data-acquisition. 

\begin{figure}%[htbp!]
    \centering
    \includegraphics[width=\textwidth]{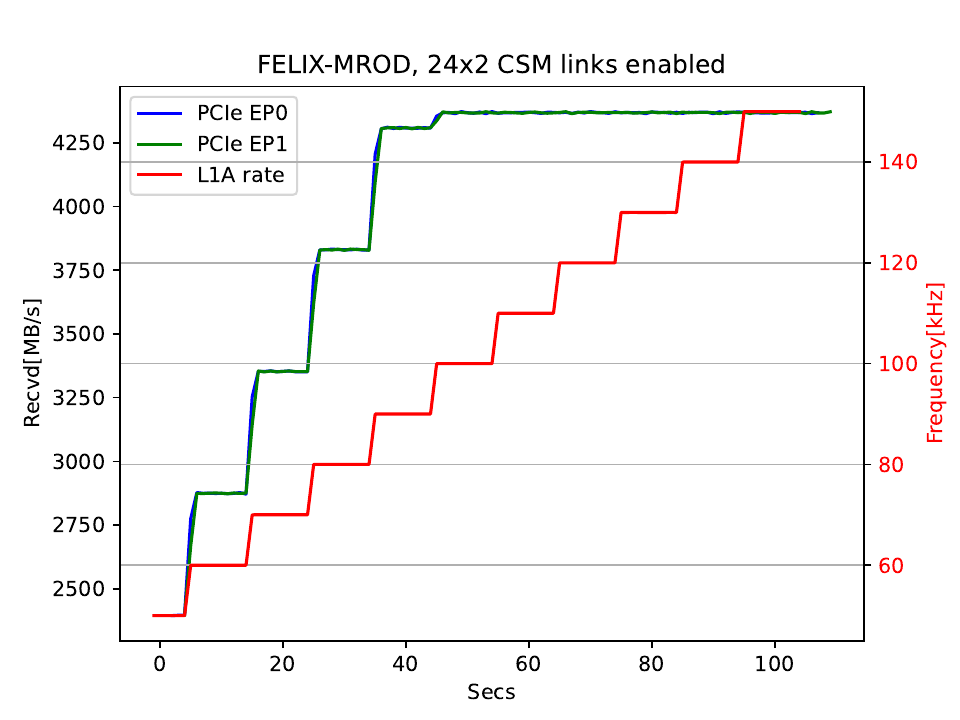}
    \caption{Measurement of the throughput, separately for each of the two PCIe endpoints of a FELIX-MROD card, for increasing trigger rates.}
    \label{fig:ScanPlot}
\end{figure}

In Table~\ref{tab:MRODrates} the computed maximum data rate for a set of benchmark scenarios is presented in which the input data of 1, 4 or 8 MROD modules is handled corresponding to the case where 6, 24 or 48 input channels of the FLX-712 are enabled, respectively. The data rate was calculated using:
\begin{equation} \text{Data rate per card} = (\text{word rate per link} \times \text{Nr. links per card}) \times \text{size of a word} \label{eq:DataRate}  \end{equation}
where the size of a word is 4 bytes and the value of the CSM clock (50 MHz) is taken as the word rate per link.
The result of the measurement with all 48 channels enabled is reported in Figure~\ref{fig:ScanPlot}, where it is shown that at 100 kHz L1A rate the data rate saturates at a plateau
of 8.8 GB/s (4.4 GB/s for each of the two endpoints (EP) of the PCIe card).
This value was found to be in agreement with the expected theoretical maximum data rate of 9.6 GB/s. The difference between the measured 8.8 GB/s and the expected 9.6 GB/s is due to the presence of idle words in-between cycles. 
In order to take this effect into account, a factor of $\frac{19}{21}=\frac{\text{Nr. words per cycle (incl. Separator)}}{19+2\text{ idle words}}$ should be applied.

\begin{figure}%[htbp!]
    \centering
    \includegraphics[width=0.75\textwidth]{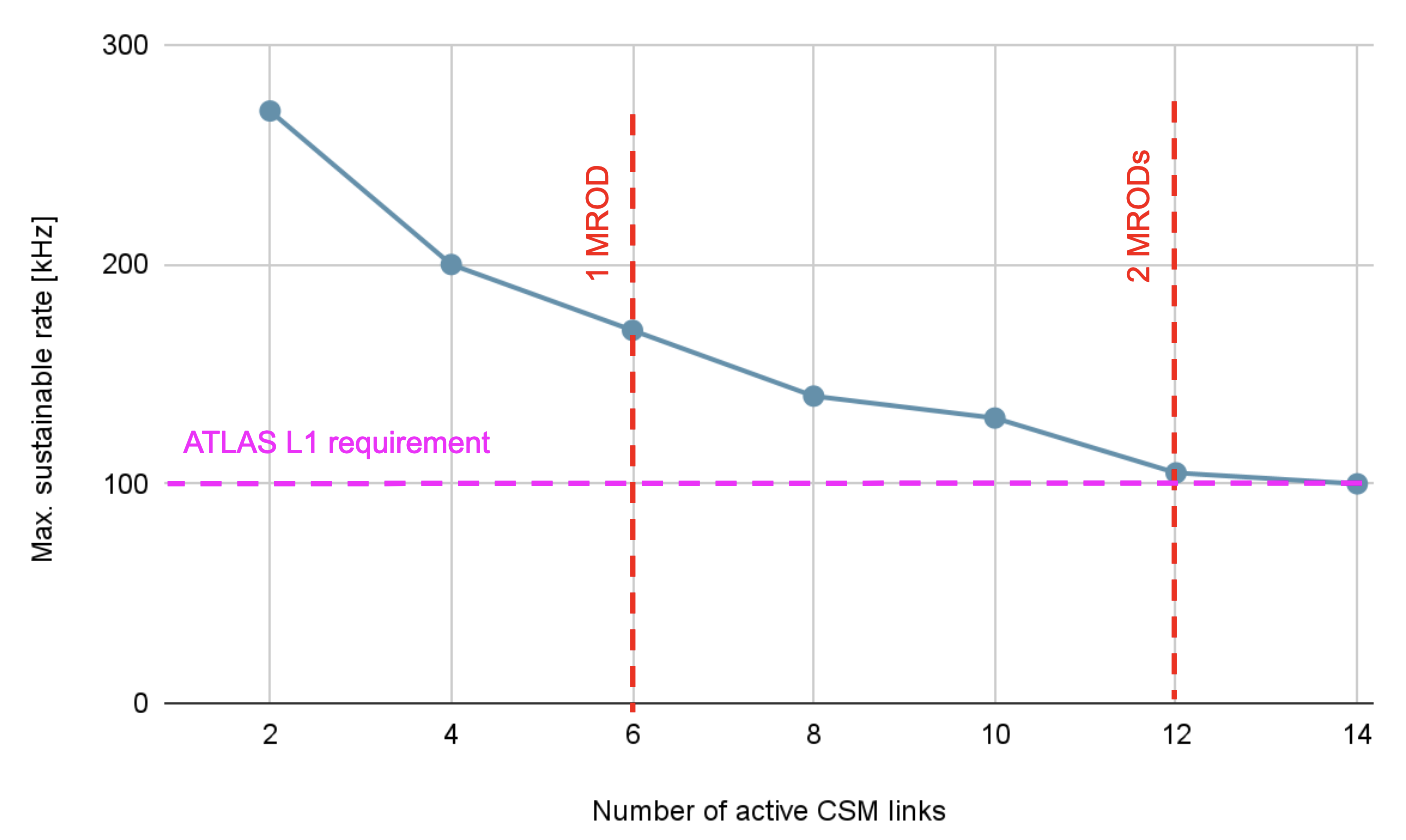}
    \caption{Maximum sustainable rate as a function of the number of active CSM links.}
    \label{fig:csmswrod_maxrate}
\end{figure}

\begin{figure}%[htbp!]
    \centering
    \includegraphics[width=\textwidth]{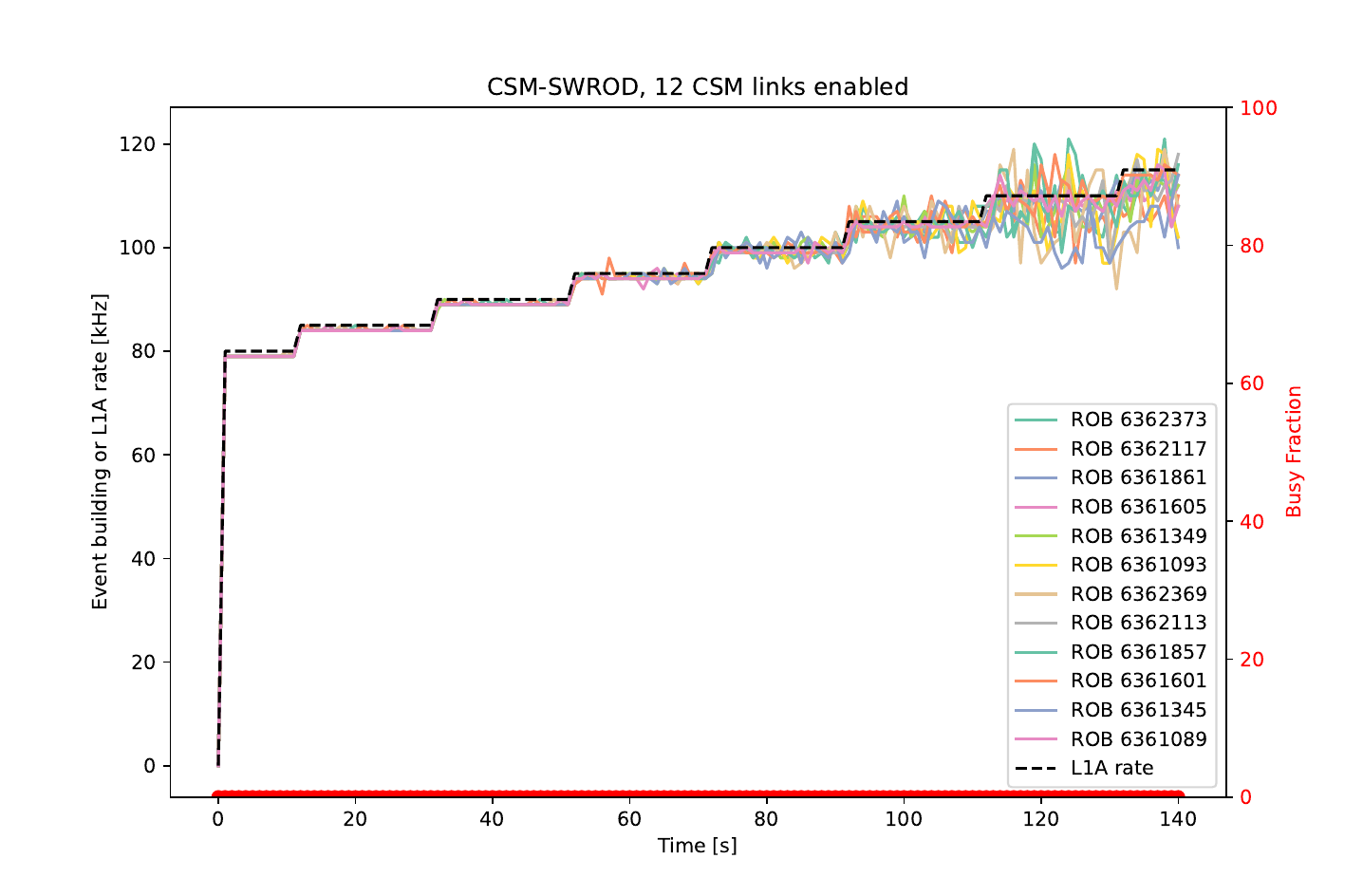}
    \caption{Measurement of the event building rate, separately for each of the 12 ROBs (one per link), for increasing trigger rates.}
    \label{fig:ROBrates}
\end{figure}

\begin{figure}[htbp]
    \centering
    \begin{subfigure}[b]{0.49\textwidth}
        \centering
        \includegraphics[width=\textwidth]{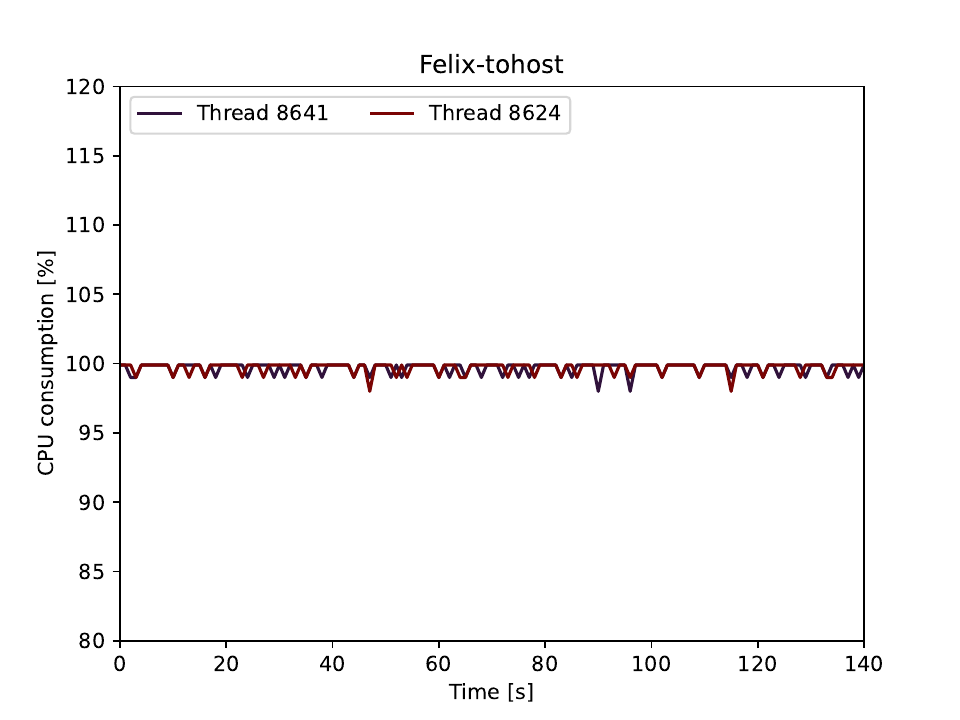}
        \caption{FELIX software application running in the FELIX server.}
        \label{fig:FelixToHost}
    \end{subfigure}
    \hfill
    \begin{subfigure}[b]{0.49\textwidth}
        \centering
        \includegraphics[width=\textwidth]{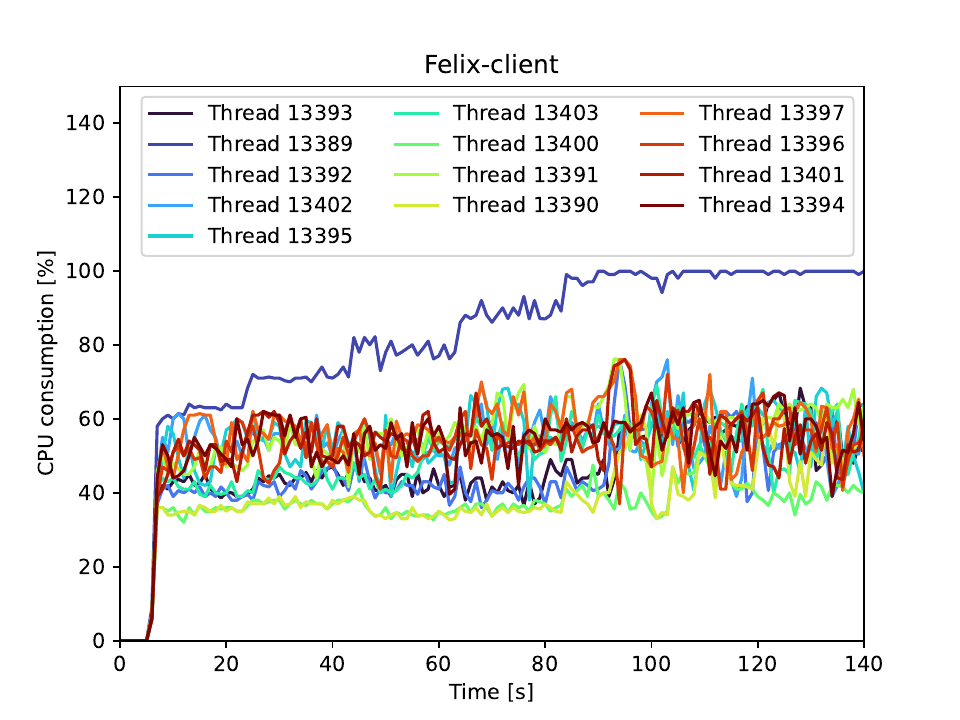}
        \caption{FELIX client threads running in the SWROD server.}
        \label{fig:FelixClient}
    \end{subfigure}
    \\ % New row
    \begin{subfigure}[b]{0.49\textwidth}
        \centering
        \includegraphics[width=\textwidth]{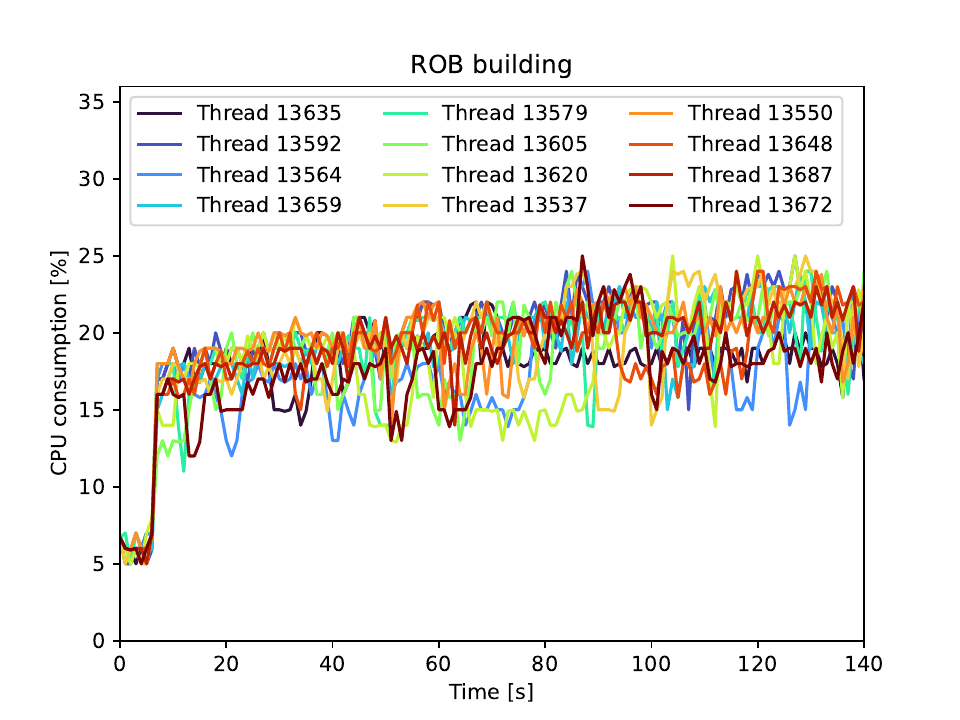}
        \caption{ROB building threads running in the SWROD server.}
        \label{fig:ROBbuild}
    \end{subfigure}
    \hfill
    \begin{subfigure}[b]{0.49\textwidth}
        \centering
        \includegraphics[width=\textwidth]{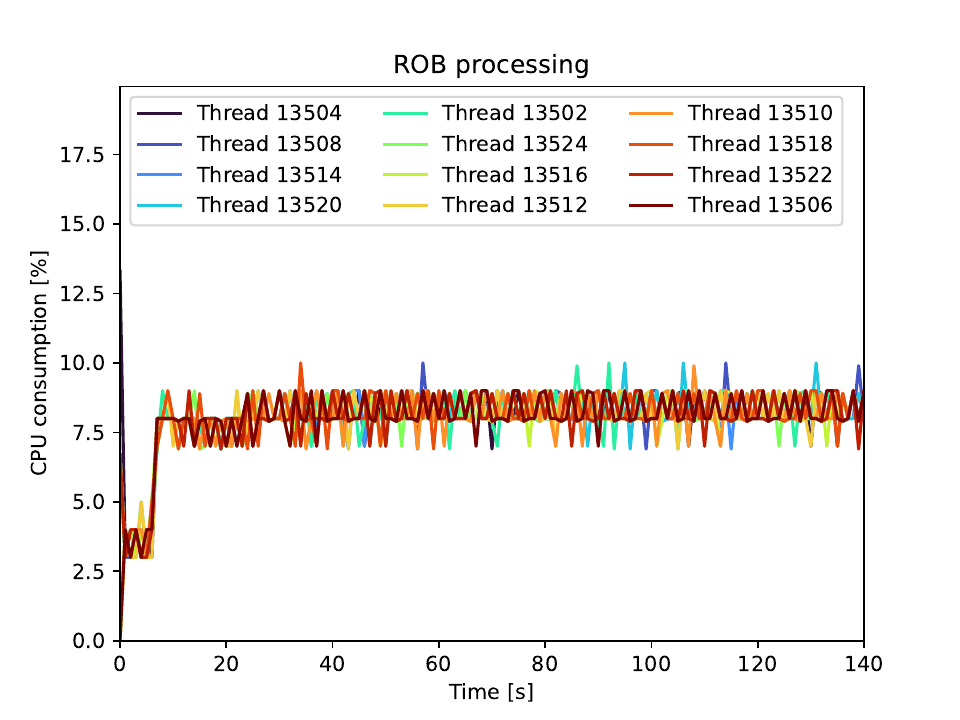}
        \caption{ROB processing threads running in the SWROD server.}
        \label{fig:ROBproc}
    \end{subfigure}
    \caption{CPU consumption over time for all the various applications involved in the test.}
        \label{fig:csmswrod_cpu}
    \end{figure}

To identify the real limit of the whole FELIX+CSM-SWROD system, multiple tests were performed, each time enabling more data links.
The goal of this stress test was to find the maximum number of CSM links (GOLs) that the system can handle without compromising the performance.
A summary of the results is given in Figure~\ref{fig:csmswrod_maxrate}.
Each point in the graph represents a test that is run with a given link configuration: last values before the system breaks (i.e. event building rate goes to 0 OR busy is raised OR errors start to appear) are recorded.
The number of links is always equally distributed across the two endpoints: ``2 links'' means 1 link to EP 0 and 1 link to EP 1.
We can conclude that the system can successfully replace 2 MROD modules, while still fulfilling the requirements of the ATLAS L1 trigger in Run 3.
For configurations with more than 14 CSM links, the performance starts to degrade to a level where the event building rate falls below the 100 kHz threshold.\\
Results of the test with 12 active CSM links are given in Figure~\ref{fig:ROBrates}. In order to guarantee optimal performance, CSM-SWROD was configured to have one ROB per link. The event building rate recorded for every ROB is on average consistent with the increasing input L1A rate. At about 80 kHz, the rates are very stable while as the frequency goes up fluctuations start to appear and become more and more important. At 100 kHz they are of the order of 5 \%.
During the test, the CPU consumption of all applications at play was monitored, as shown in Figure~\ref{fig:csmswrod_cpu}.
Felix-tohost threads, one for each of the two EPs used, saturate at 100 \%.
There are in total 13 Felix-client threads: one per ROB plus one handling the L1A signal (Thread 13389), which is the one reaching 100 \% CPU during the high frequency (higher than 100 kHz) steps.
ROB building and ROB processing threads take only a relatively minor fraction of the CPU and don't represent a bottleneck for the system in this configuration.


\begin{table}
\caption{FELIX-MROD card output rates for three benchmark scenarios.}
\centering
\begin{tabular}{cccc}
\hline
%\toprule
& CSM links per card & Word rate per card (Mwords/s) & Data rate per card (GB/s)\\
\hline
%\midrule
1 MROD & 6 & 300 & 1.2 \\
4 MROD & 24 & 1200 & 4.8 \\
8 MROD & 48 & 2400 & 9.6 \\
\hline
%\bottomrule
\end{tabular}
\label{tab:MRODrates}
\end{table}


\subsection{Tests with standalone MDT chambers}
\subsubsection{Tests at CERN's BB5 facility}
\begin{figure}[htbp]
\centering
%\subfloat{\includegraphics[width=0.58\textwidth]{figures/bb5_mdt.png}}
\includegraphics[width=0.58\textwidth]{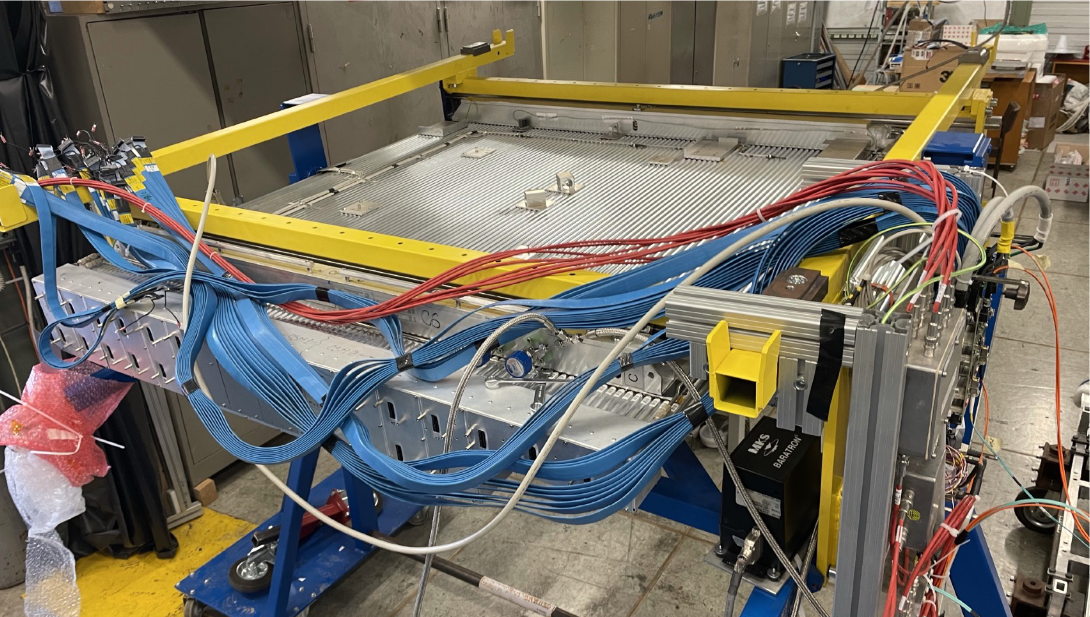}
\hspace{0.5cm}
%\subfloat{\includegraphics[width=0.35\textwidth]{figures/bb5_csm.png}}
\includegraphics[width=0.35\textwidth]{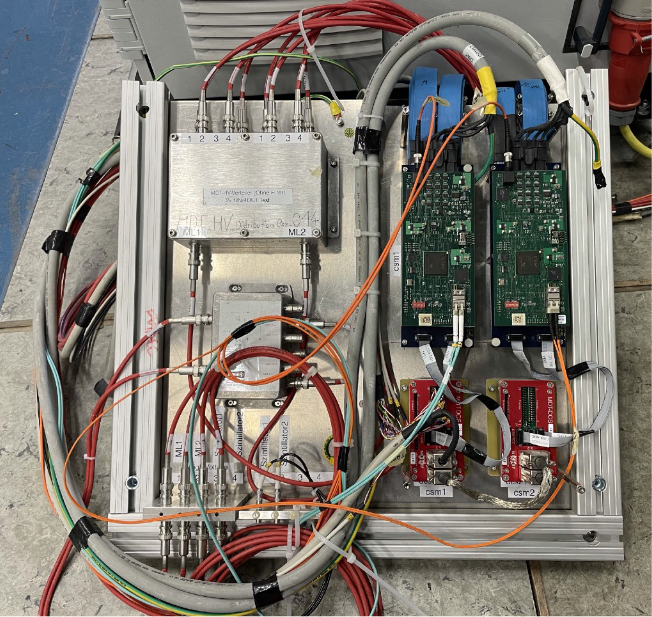}
\caption{Pictures of the test MDT chambers (a) and CSM and MDT-DCS modules (b)
in the BB5 facility at CERN.}
\label{fig:bb5}
\end{figure}

The first integration test with MDT chambers has been performed in the BB5
facility at CERN in 2021. 
The setup, shown in Figure~\ref{fig:bb5}, comprised:
\begin{itemize}
\item two MDT chambers with 18 and 13 TDCs respectively. Each chamber was
equipped with a CSM and an MDT-DCS module (MDM) for chamber front-end electronics configuration,
\item a TTC system,
\item a server interfaced via CAN-BUS to the DCS system,
\item a FELIX-MROD server equipped with a 48-channel FLX-712 and a 10 GbE Mellanox
Connect-X3 network interface.
\end{itemize}
For the duration of the test the MDT tubes were filled with gas and connected to
a high voltage power supply. Since the MDTs were active and the BB5 facility on
surface, the source of hits were cosmic rays and noise. Hits were not used to
trigger data acquisition; trigger signals were produced independently up to a rate of \SI{100}{kHz}.
This operation mode was particularly suitable for testing FELIX-MROD.
In response to L1As the TDCs produced data consisting of header and trailer words and of time stamps due to any hits occurring in the time windows associated with the L1As. Since FELIX-MROD is a data acquisition system, and does not
perform any event reconstruction, the cause of the hits is irrelevant.

\begin{figure}[htbp]
    \centering
    \includegraphics[width=0.75\textwidth]{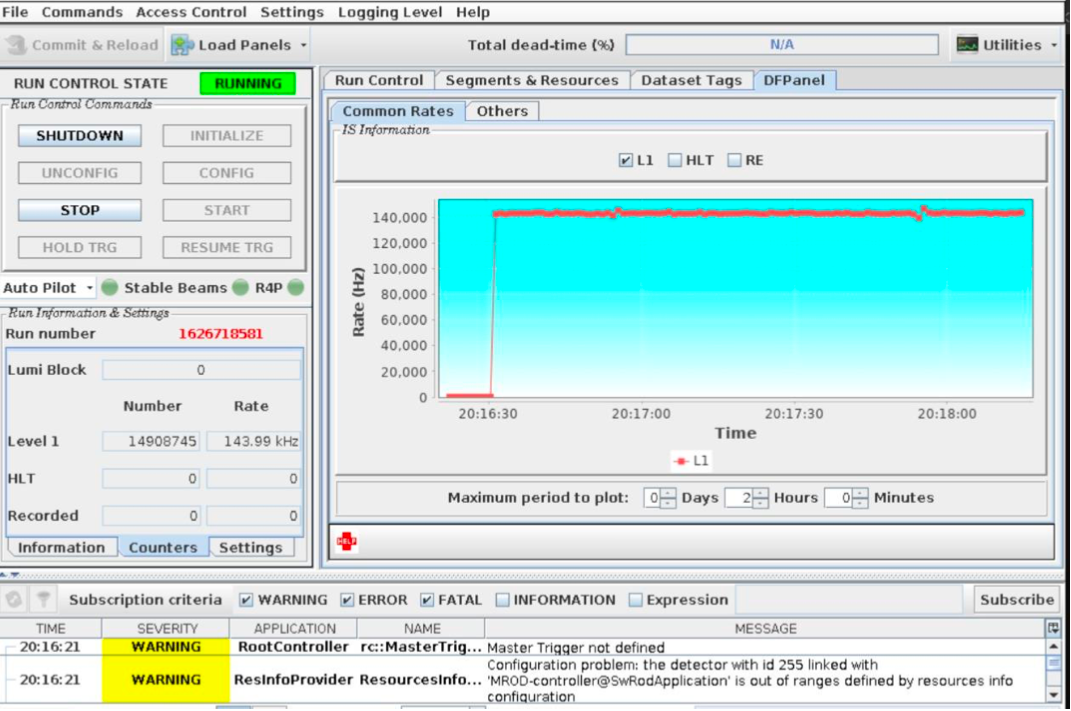}
    \caption{Fragment building rate of CSM-SWROD as displayed in the
    ATLAS TDAQ GUI. Fragments are built in TTC-driven mode reading two CSMs with 18
    and 13 TDCs at L1A rate of \SI{143}{kHz}}
    \label{fig:bb5_igui}
\end{figure}

\begin{figure}[htbp]
    \centering
    \includegraphics[width=\textwidth]{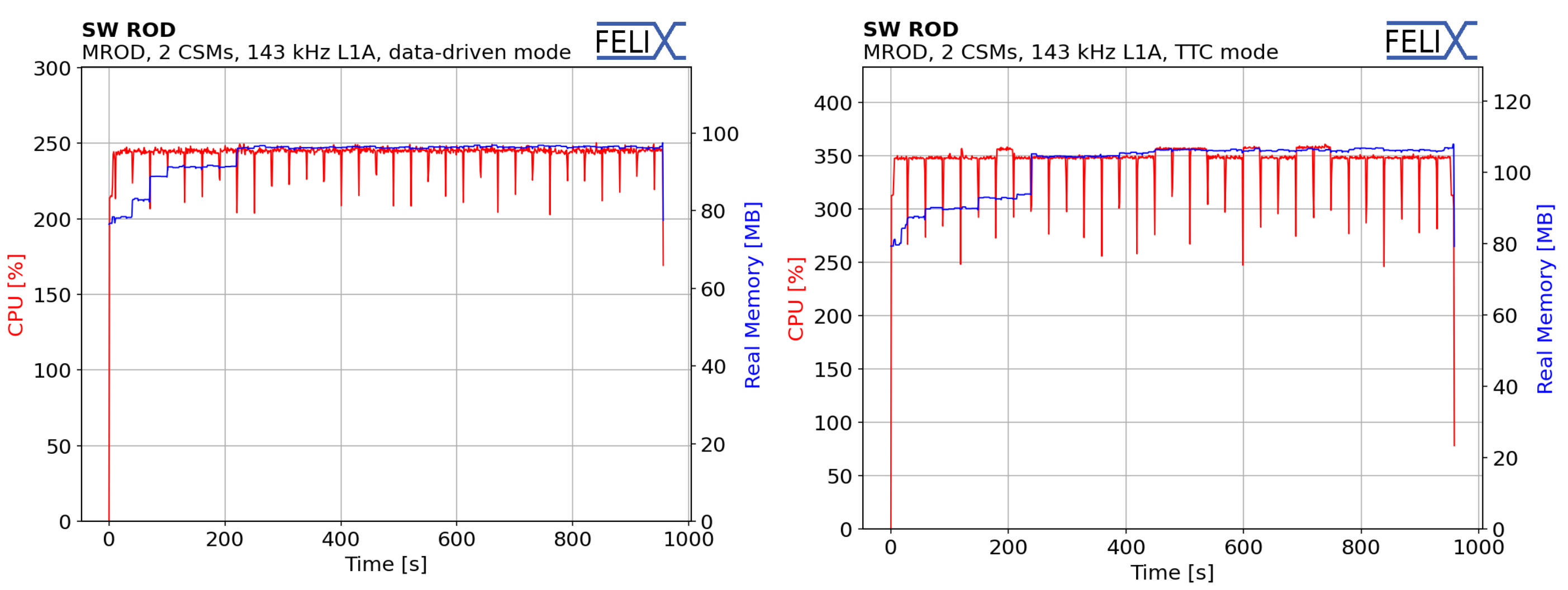}
    \caption{CPU and memory use of CSM-SWROD while reading two CSMs with 18
    and 13 TDCs at L1A rate of \SI{143}{kHz} in both data-driven and TTC-driven
    mode.}
    \label{fig:bb5_cpu}
\end{figure}

No firmware issues were encountered and link alignment was immediate. 
FELIX-MROD read out two CSMs at a
L1A rate of \SI{143}{kHz} (the excess over \SI{100}{kHz} was due to the 
imprecise calibration of the trigger generator). CSM-SWROD has been run in two modes:
``data-driven'', in which L1IDs are read exclusively from data messages, and
``TTC-driven'' in which L1IDs are obtained from the TTC link and searched for in the data. 
In both modes, stable operation has been observed as shown in the screen capture of Figure~\ref{fig:bb5_igui}. Data fragments produced by CSM-SWROD were
dumped to a local disk for analysis, limiting the duration
of data-taking periods. 
No data corruption has been observed in the recorded data.
The performance of CSM-SWROD is shown in Figure~\ref{fig:bb5_cpu}: the memory usage
reaches a plateau at 1 GB while 3.5 cores are used to process data from 2 CSMs.
However not all CPU resources are used for data processing in CSM-SWROD; a part is
used for data reception by the underlying network library netio-next.

\subsubsection{Reliability tests at LMU}
\begin{figure}
    \centering
    \includegraphics[width=0.75\textwidth]{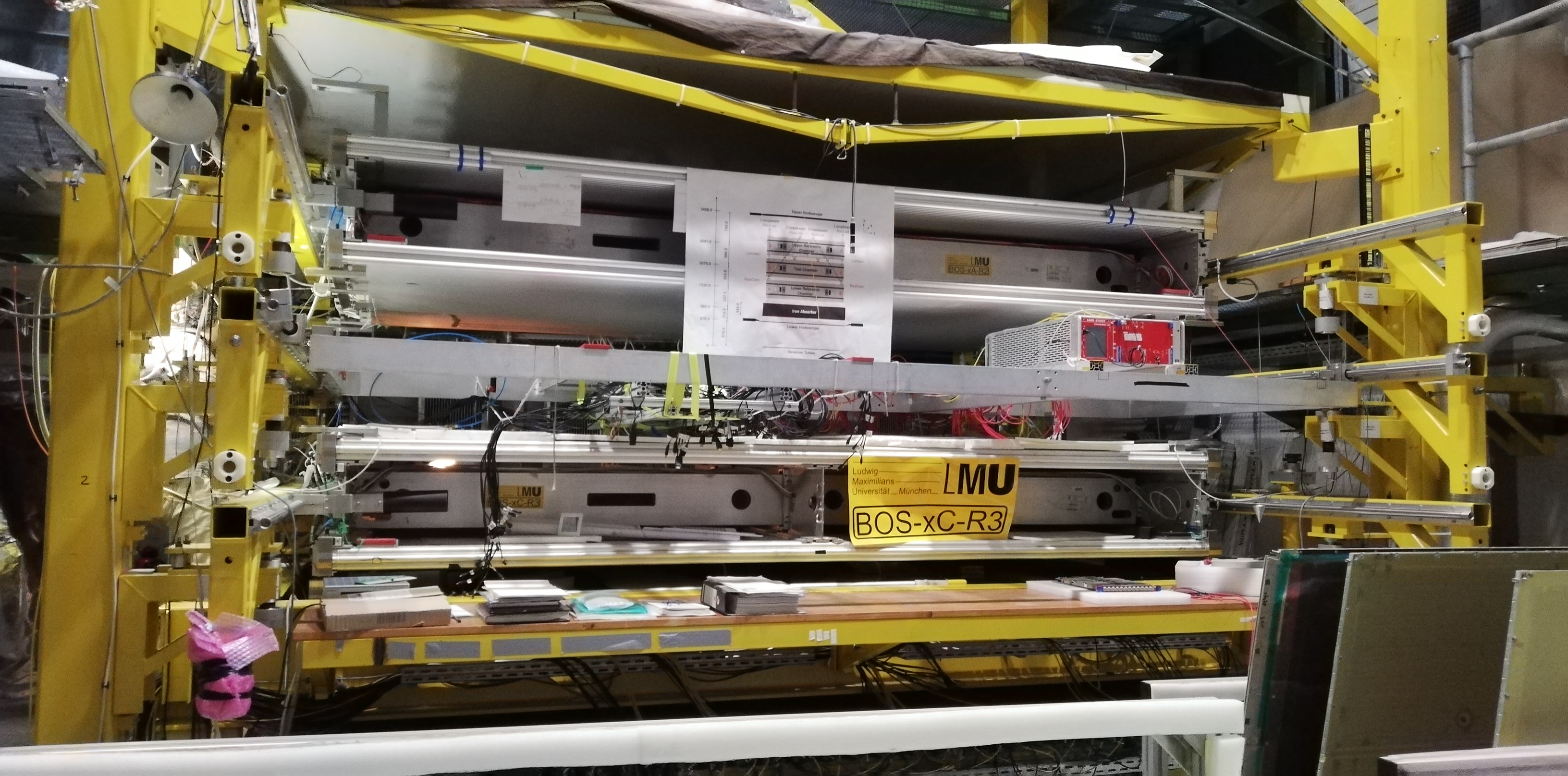}
    \caption{Picture of the MDT chambers at the CRF in Garching. Two layer of scintillators are positioned above and below both chambers.}
    \label{fig:GarchingMDT}
\end{figure}

\begin{figure}
    \centering
    \includegraphics[width=0.75\textwidth]{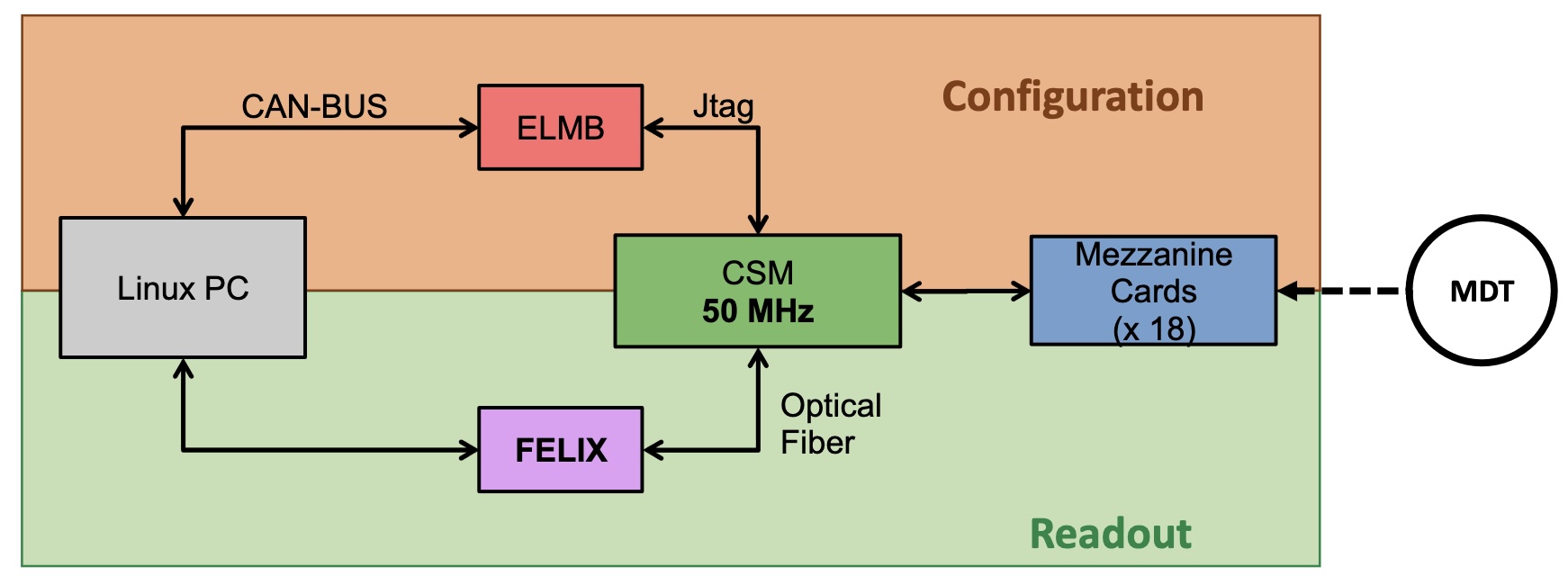}
    \caption{Schematic representation of the Configuration \& Readout architecture of the MDT chambers at the CRF in Garching.}
    \label{fig:GarchingDAQ}
\end{figure}

\begin{figure}
    \centering
    \includegraphics[width=0.75\textwidth]{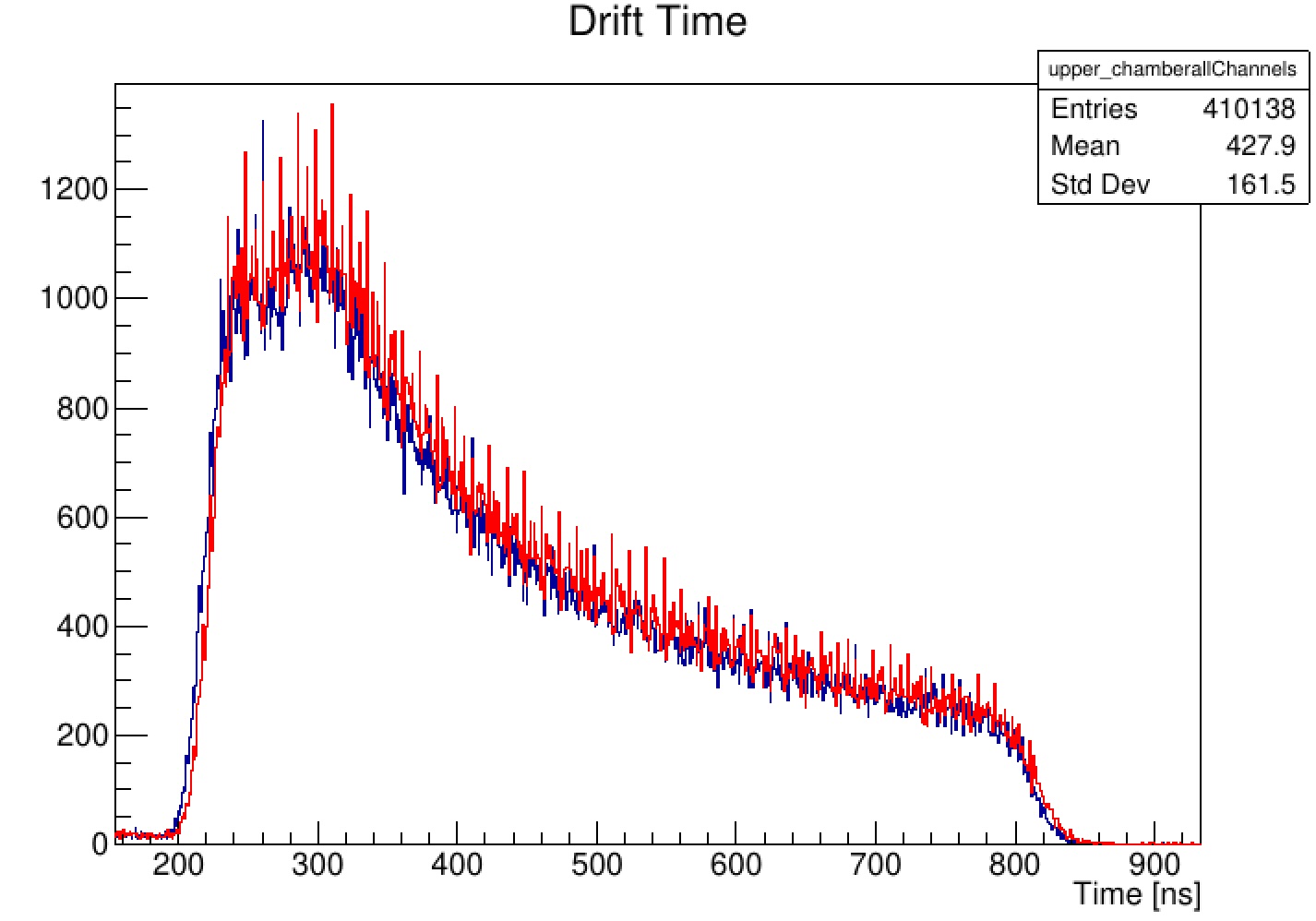}
    \caption{Drift time spectra for the two MDTs separately.}
    \label{fig:MuonDriftSpectra}
\end{figure}

\begin{figure}
    \centering
    \includegraphics[width=0.75\textwidth]{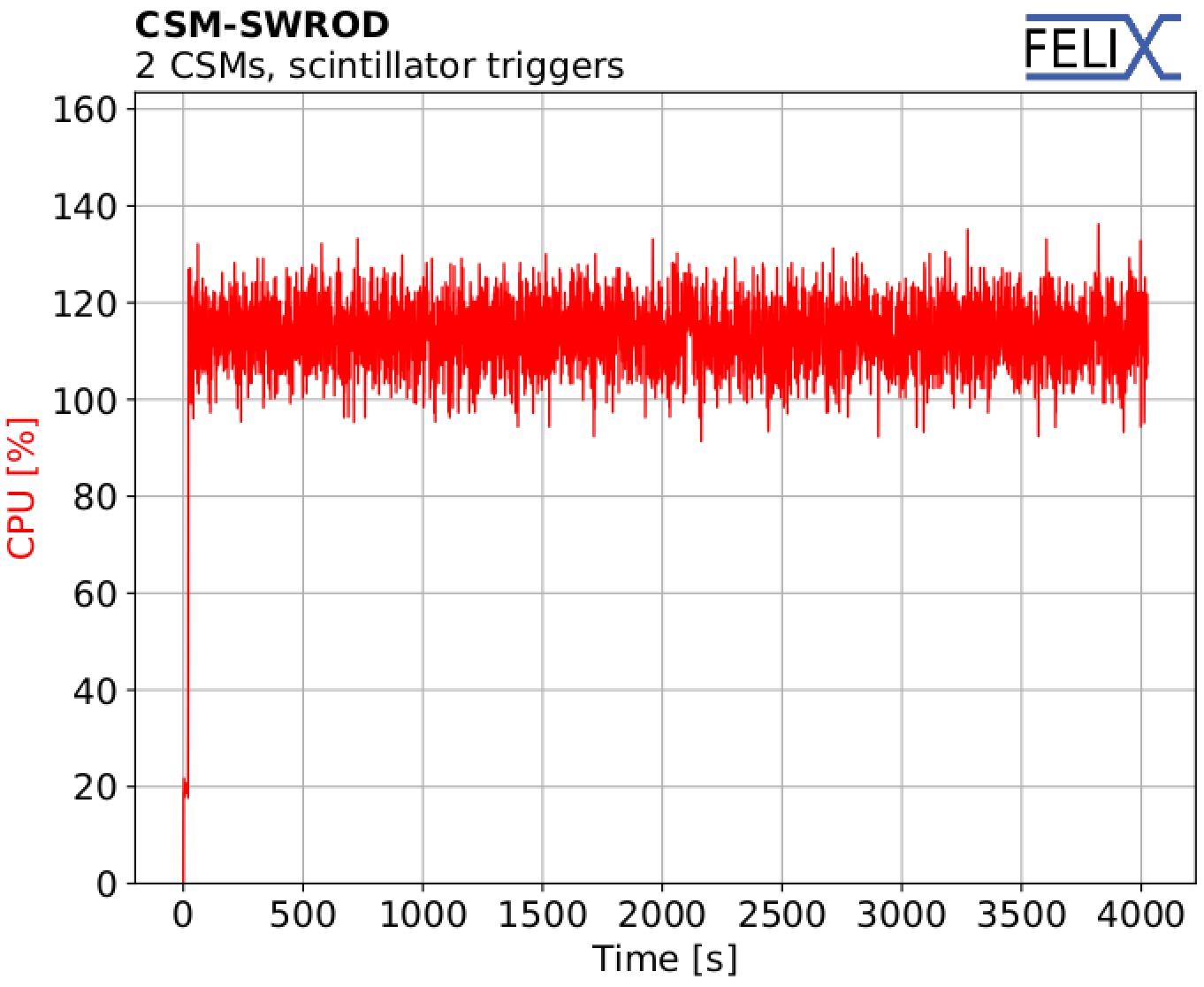}
    \caption{CPU usage of CSM-SWROD as a function of time.}
    \label{fig:GarchingCPU}
\end{figure}

FELIX-MROD may be adopted by the ATLAS group at the Ludwig Maximilian University (LMU) in Munich in view of testing the Phase-II Upgrade of the ATLAS Muon Spectrometer~\cite{florianThesis}.
At the Cosmic Ray Facility (CRF) in Garching there are two fully functional MDT chambers, each one equipped with a 50 MHz CSM,
as well as two sets of scintillators used for triggering on cosmic muons. A picture is given in Figure~\ref{fig:GarchingMDT}.
As shown in Figure~\ref{fig:GarchingDAQ}, the CSMs are configured via MDT-DCS modules (ELMB in the figure), which are controlled by a Linux server. 
The same server hosts one FLX-712 card, loaded with FELIX-MROD firmware, reading out both CSMs simultaneously and runs CSM-SWROD to process the incoming data.
This setup has been used to test the stability and reliability of FELIX-MROD in data-taking conditions similar to those of a cosmic run.
Given the low trigger rate of the scintillators (about 100 Hz), the system was not meant to be particularly under stress.
Nonetheless, this mode of operation allowed to run over a long period of time without risking to fill up the limited available storage space.
Additionally, recorded data was reconstructed and analyzed by means of LMU's custom scripts and drift time spectra as expected were obtained, as shown in Figure~\ref{fig:MuonDriftSpectra}. These spectra are not corrected for the 25 ns time jitter coming from the fact that cosmic muons are not correlated with the 40MHz clock.
The CPU consumption of CSM-SWROD was monitored for about one hour as shown in Figure~\ref{fig:GarchingCPU}.
For the duration of the tests, the application ran successfully without errors.


\section{Test with ATLAS}
\label{sec:integration}
% \subsection{ATLAS MDT}\label{subsec:atlas}
% \subsection{Munich MDT}\label{subsec:lmu}
One FELIX-MROD server has been installed in the ATLAS counting room (USA15) in 2021.
The 6 inputs of an MROD were provided
to FELIX-MROD by passive optical splitting. The six inputs received data from MDTs in the barrel. The optical power measured by
the FLX-712 was sufficient for 5 out of 6 CSMs. Details are reported in Table~\ref{tab:csm_atlas}.

Aiming to include FELIX-MROD in the combined ATLAS partition, CSM-SWROD had
been integrated in the Muon segment. However, misconfiguration of the HLT
request handler and TDC masks did not allow to run successfully during the 1
hour slot assigned on 26 October 2021.
Nevertheless data samples were recorded with low-level FELIX tools and CSM-SWROD
during the pilot beam run of October 28-29 and found to be consistent with other MDT data.

\begin{table}
\centering
\begin{tabular}{|c|c|c|}
%###\toprule
\hline
Channel & Nr. of TDCs & optical power [$\mu W$]\\
%###\midrule
\hline
BIL1C09 & 10 & 121 \\
BIL2C09 & 12 & 142 \\
BML1C09 & 10 & 140 \\
BML2C09 & 14 & 93 \\
BOL1C09 & 14 & 45 \\
BOL2C09 & 18 & 89 \\
%###\bottomrule
\hline
\end{tabular}
\caption{Identifiers of the six ATLAS CSMs connected to FELIX-MROD. The optical
power is measured by FLX-712 after the optical splitter.}
\label{tab:csm_atlas}
\end{table}

\section{Conclusions} \label{sec:conclusion}
The FELIX-MROD project has been developed with
the aim of overcoming a possible massive failure of
the legacy MROD modules.
Not only it can successfully perform the same
functionality of the MROD (and ROS) in the ATLAS
MDT readout architecture, but, due to the
versatility of the FELIX system, it is also a more
efficient solution: while a single MROD module can receive the data streams from six CSMs, the hardware design of an
FLX-712 card allows to accommodate up to 48 channels.
When taking into account the performance requirements of the ATLAS readout system for Run 3, a single FELIX-MROD unit consisting of FELIX server with FLX-712 PCIe card and a server running the CSM-SWROD application can successfully replace two MROD modules. However, this number most likely can be increased by upgrading the FELIX-MROD firmware so that 8 to-host DMA channels, 4 per endpoint, are implemented (the latest standard FELIX firmware has, depending on the type of flavor, 4 or 5 DMA to-host channels per endpoint). In that case 8 felix-tohost instances could run in parallel on different cores of the CPU of the FELIX server. As it has been found that one instance can handle the data from 6 CSM links and two instances the data from 12 links, the expectation is that 8 instances can handle the data from 48 CSM links. Their data could be sent to 4 CSM-SWROD instances, either running on different servers, or running in parallel on e.g. a single 64-core CPU server, so that in total 8 MROD modules could be replaced.
The use of FELIX-MROD for testing of sMDT
chambers produced for the Phase-II ATLAS upgrade
is considered and integration studies in this
direction are currently ongoing.

% We discourage the use of inline figures (e.g. \texttt{wrapfigure}), as they may be
% difficult to position if the page layout changes.

% We suggest not to abbreviate: ``section'', ``appendix'', ``figure''
% and ``table'', but ``eq.'' and ``ref.'' are welcome. Also, please do
% not use \texttt{\textbackslash emph} or \texttt{\textbackslash it} for
% latin abbreviaitons: i.e., et al., e.g., vs., etc.

%%%------------------------%%%
% \appendix
% \section{Some title}
% Please always give a title also for appendices.
%%%------------------------%%%

%\acknowledgments

%This is the most common positions for acknowledgments. A macro is
% available to maintain the same layout and spelling of the heading.

%\paragraph{Note added.} This is also a good position for notes added
% after the paper has been written.

% Bibliography

%% [A] Recommended: using JHEP.bst file
\bibliographystyle{JHEP}
\bibliography{biblio.bib}

%% or
%% [B] Manual formatting (see below)
%% (i) We suggest to always provide author, title and journal data or doi:
%% in short all the informations that clearly identify a document.
%% (ii) please avoid comments such as "For a review'', "For some examples",
%% "and references therein" or move them in the text. In general, please leave only references in the bibliography and move all
%% accessory text in footnotes.
%% (iii) Also, please have only one work for each \bibitem.

\end{document}